\documentclass{article}
\usepackage{url}
\usepackage{xcolor}
\usepackage[colorlinks=true,
	linkcolor=violet,
	citecolor=black]{hyperref}

\usepackage{amssymb, amsfonts, amsmath}
\usepackage{graphicx, natbib}
\usepackage{algorithm2e}

\newcommand{\BIT}{\begin{itemize}}
\newcommand{\EIT}{\end{itemize}}
\newcommand{\BNUM}{\begin{enumerate}}
\newcommand{\ENUM}{\end{enumerate}}

\def\reals{\mathbb{R}} 
\def\naturals{\mathbb{N}} 
\def\simplex{\Delta} 
\renewcommand{\exp}[1]{\operatorname{exp}\left(#1\right)} 

\def\absarg#1{\left|#1\right|}

\def\Unif{\textnormal{Unif}}
\def\Mult{\textnormal{Mult}}

\def\Dir{\textnormal{Dir}}

\def\*#1{\mathbf{#1}}

\newcommand{\win}{w_{\text{in}}} 
\newcommand{\wout}{w_{\text{out}}} 

\usepackage{placeins}
\usepackage{float}
\usepackage{longtable}
\usepackage{caption}

\title{Multiscale Analysis of Count Data through \\ Topic Alignment}

\author{JULIA FUKUYAMA$^\ast$ \\ 
\textit{Department of Statistics, Indiana University Bloomington} \\
{jfukuyam@indiana.edu}\\
KRIS SANKARAN$^\ast$ \\
\textit{Department of Statistics, University of Wisconsin - Madison}\\
{ksankaran@wisc.edu}\\
LAURA SYMUL$^\ast$\\
\textit{Department of Statistics, Stanford University} \\
{lsymul@stanford.edu}\\
}

\begin{document}
\maketitle
\footnotetext{All authors contributed equally to this work and should be included in correspondence.}

\begin{abstract} {
Topic modeling is a popular method used to describe biological count data.  With topic models, the user must specify the number of topics $K$. Since there is no definitive way to choose $K$ and since a true value  might not exist, we develop a method, which we call {\em topic alignment}, to study the relationships across models with different $K$. In addition, we present three diagnostics based on the alignment. These techniques can show how many topics are consistently present across different models, if a topic is only transiently present, or if a topic splits in more topics when $K$ increases. This strategy gives more insight into the process generating the data than choosing a single value of $K$ would.
We design a visual representation of these cross-model relationships, show the effectiveness of these tools for interpreting the topics on simulated and real data, and release an accompanying R package, \href{https://lasy.github.io/alto}{\texttt{alto}}. }
{topic model; microbiota; community analysis; multiresolution; mixed membership models}
\end{abstract}

\section{Introduction}

Topic models are probabilistic models for dimensionality reduction of count data \citep{blei2003latent}. They are widely used in modern biostatistics, finding application in population genetics, genome-wide association studies, metabolomics, and microbiota studies \citep{al2019inference, reder2021supervised, leite2020you, gonzalez2019cistopic, sankaran2019latent}.

These models are appealing because they are more expressive than clustering yet have simple interpretations \citep{airoldi2014introduction}. Like clustering, topic models provide a small set of ``prototypical'' data points; this enables summarization of the overall collection. Unlike clustering, where each sample must belong to exactly one cluster, topic models support varying grades of membership. Therefore, samples are allowed to smoothly blend from one prototype to another. Alternatively, topic models can be viewed as a form of constrained dimensionality reduction, where factors and loadings are constrained to lie on the probability simplex \citep{carbonetto2021non}. The sum-to-one constraint can make the results more interpretable than standard PCA, NMF, or factor analysis: each sample can be written as a mixture of underlying types, and each topic is a probability distribution across data dimensions. For example, for microbiota data, each topic can be interpreted as a sub-community of bacteria and each sample is a mixture of a few underlying sub-communities.

\begin{figure}
    \centering
    \includegraphics[width=\textwidth]{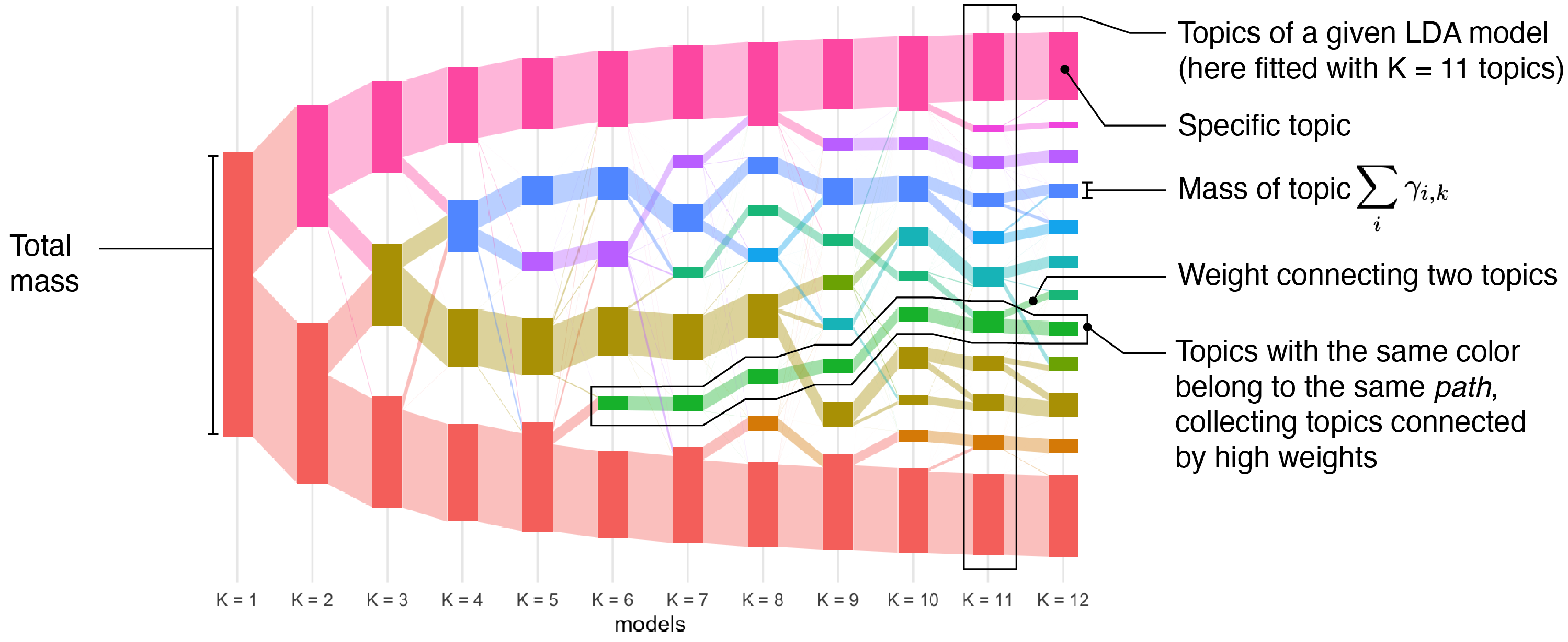}
    \caption{How to read a topic alignment. Construction of weights is discussed in Section \ref{sec:alignment} and paths are defined in Subsection \ref{subsec:paths}.}
    \label{fig:annotated_alignment}
\end{figure}

Like most clustering and dimensionality reduction methods, topic models come with a hyperparameter, $K$, that controls the complexity of the resulting fit, and choosing a good value of $K$ to aid downstream analysis remains a challenge. Past work has focused on automatic selection of this hyperparameter, typically by referring to the marginal likelihood of a test set \citep{wallach2009evaluation, kass1995bayes}. In this study, we explore an alternative, a process we call \emph{topic alignment} (Figure \ref{fig:annotated_alignment}), which is based on describing how models fit across a range of $K$ relate to one another.

This reframing has appeared in previous literature, though typically in the context of new models, rather than new algorithms applied to existing models. For example, a hierarchical extension of topic models \citep{blei2003hierarchical} provides a similar multiscale interpretation of topic structure. However, computational challenges have made these models somewhat difficult to extend and apply, compared to fixed $K$ topic models. In the hierarchical clustering context, a comparison across choices of $K$ is central to the HOPACH algorithm \citep{pollard2005cluster}, which evaluates cluster stability using a bootstrap procedure. 

Instead of introducing a novel multiscale model, we focus on post-estimation comparison of an existing ensemble. This is in the spirit of methodology for comparing clusterings \citep{meilua2007comparing, wagner2007comparing}, which introduce metrics for navigating the space of clustering results. Similarly, a description of the relationship between models across choices of $K$ is provided by graphical posterior predictive analysis \citep{gelman2004exploratory, gelman2013philosophy}. A posterior predictive check can highlight the lack of fit at particular choices of $K$, in addition to guiding the selection of $K$. We also note a connection to Tukey’s process of iterative data structuration \citep{tukey1977exploratory, holmes1993comment, holmes2018modern}. 

Alignment of models across scales naturally supports a coarse-to-fine analysis, ensuring that subtle patterns can be related to their overall context.
First, this helps navigate the interpretability-expressivity trade-off associated with different choices of $K$. Models with small $K$ tend to be more interpretable, but may suppress interesting variation in the data. Conversely, models with large $K$ are more faithful to the data, but can be overwhelming to the analyst. By streamlining comparison across $K$, we get the best of both worlds — topics at large values of $K$ can be interpreted in context of the coarser ones to which they relate. Second, topic alignment is still relevant to the challenge of choosing $K$. In a way that is made precise in Section \ref{sec:diagnostics}, true topics tend to be more stable across choices of $K$, while spurious ones are more transient.
Finally, alignment can help practitioners discover mis-specifications in topic models. For example, it is biologically plausible that microbiota data deviate from the topic model generative mechanism in the following ways:
\begin{itemize}
\item Elevated heterogeneity: Topic models assume that all samples are a mixture of a few underlying sub-communities. If samples have more heterogeneity than expected — e.g., due to unmodeled external factors — then topic models may be inappropriate, even for large $K$.
\item Strain switching: There may be strains of a species that compete for the same ecological niche. If one strain is successful, then the other would be expected to be absent. This can result in sharp differences in strains within an otherwise well-defined community structure.
\end{itemize}
In Section \ref{sec:simulations}, we generate data inspired by these phenomena and apply topic alignment to them. We describe the degree to which the resulting alignments reflect underlying heterogeneity or switching. As long as the mis-specification is not too subtle, topic alignment can suggest specific structure to incorporate into follow-up analysis.

In the remainder of this paper, we present the following contributions:
\begin{itemize}
\item The design of algorithms and diagnostics to support the comparison of topic models fit across a range of scales $K$.
\item An analysis of the properties of these algorithms and diagnostics, using simulation experiments across several generative mechanisms.
\item An illustration of topic alignment applied to a microbiota data analysis problem.
\item The release of an R package, \texttt{alto}, implementing these methods.
\end{itemize}

Sections \ref{sec:background} and \ref{sec:methods} review relevant background material and present algorithms and diagnostics for topic alignment, respectively. Subsection \ref{subsec:package} briefly describes the \texttt{alto} package and the workflow that it supports. Section \ref{sec:simulations} presents a suite of simulation experiments, with an emphasis on exploring model mis-specification through alignment. Section \ref{sec:analysis} describes the application of topic alignment to a data analysis problem associated with the vaginal microbiota. This is a setting where high-level structure is dominated by a few well-known species, but where additional, systematic variation is present at finer scale.

\section{Background}
\label{sec:background}

We first review topic models. Then, we summarize approaches to compare probability distributions, which are used in Section \ref{sec:methods}.

\subsection{Latent Dirichlet Allocation}

Latent Dirichlet Allocation (LDA) is a flexible way to summarize high-dimensional count data \citep{blei2003latent}. Suppose that the data are made up of $N$ samples $x_{i} \in \naturals^{D}$. For example, in text analysis, these are the counts of $D$ words across $N$ documents\footnote{A table of all notation is given in the supplementary materials.}. In the data analysis given in Section \ref{sec:analysis}, these are the counts of $D$ Amplicon Sequence Variants (ASVs)\footnote{This is the number of times specific regions of the 16S rRNA gene have been sequenced -- see \cite{callahan_replication_2017} for details of 16S sequencing technology.} across $N$ samples collected from the study participants. Let $n_{i} = \sum_{d}x_{id}$ be the total count of sample $i$. Then, LDA supposes that each $x_{i}$ is drawn independently according to
\begin{align*}
x_i \vert \gamma_i &\sim \Mult\left(n_{i}, B\gamma_{i}\right) \\
\gamma_{i} &\sim \Dir\left(\lambda_{\gamma} \cdot \*1_{K}\right),
\end{align*}
where the $K$ columns $\beta_{k}$ of $B \in \simplex^{D}$ lie in the $D$ dimensional simplex and are themselves drawn independently from
\begin{align*}
\beta_{k} \sim \Dir\left(\lambda_{\beta}\cdot \*1_{D}\right).
\end{align*}
In this mechanism, $\gamma_{i}\in \simplex^{K}$ can be interpreted as mixed-membership weights, with each $\gamma_{ik}$ giving the degree to which sample $i$ ``belongs'' to topic $K$. Since each $\gamma_{i}$ can vary continuously through the simplex, the model is more flexible than a simple clustering model, which would assign each sample to exactly one of $K$ clusters (i.e., the simplex corners). 
The three hyperparameters in this model are the number of topics $K$ and the prior parameters $\lambda_{\gamma}, \lambda_{\beta}$. Large $\lambda_{\gamma}$ and $\lambda_{\beta}$ result in Dirichlet distributions that place more mass near the uniform distribution. Small $\lambda_{\gamma}$ and $\lambda_{\beta}$ place more mass on edges and corners of the simplex, resulting in sparser $\gamma_{i}$ or $\beta_k$, respectively.

In the case of microbiota analysis, each $\beta_{k}$ corresponds to a pattern of ASV abundance. Each sample $i$ is a mixture of these underlying communities, with mixing weights  $\gamma_{i}$. Note that, though the topics are amenable to compositional interpretations — the $\beta_k$ lie on the simplex —  the original count data are modeled directly, rather than initially transformed to centered-log-ratios, for example. This makes it possible to account for differential uncertainty in samples with high and low sequencing depth and decreasing the amount of processing that takes places between raw data and final interpretation, reducing the risk for analysis errors.

\subsection{Simplex Distances and Optimal Transport}

We next review methods for comparing probability distributions. These are useful in the LDA context, because the parameters $\gamma_i$ and $\beta_k$ all lie on the probability simplex.

We first consider distances on the simplex. Let $p, q \in \simplex^{D}$ (i.e., two discrete probability distributions over $D$ categories). The Jensen-Shannon Divergence (JSD) between them is defined as 
\begin{align*}
JSD\left(p, q\right) := \frac{1}{2}\left[\text{KL}\left(p\vert\vert \frac{1}{2}\left(p + q\right)\right) + \text{KL}\left(q \vert \vert \frac{1}{2}\left(p + q\right)\right)\right],
\end{align*}
where $\text{KL}\left(a \vert\vert b\right) := \sum_{i}a_i \log\left(\frac{a_i}{b_i}\right)$ is the Kullback-Liebler divergence between $a$ and $b$. The JSD can be viewed as a symmetrized version of the Kullback-Liebler divergence, allowing it to serve as a distance measure. Intuitively, for $p$ and $q$ to have low JSD to one another, samples from either distribution should have high probability under the averaged distribution $\frac{1}{2}\left(p + q\right)$. Alternatively, the cosine similarity $\text{cossim}\left(p, q\right) := \frac{p^{T}q}{\|p\|_{2}\|q\|_{2}}$ may be used. The numerator here is large when both $p$ and $q$ place high mass on the same coordinates, and the denominator is smallest when both $p$ and $q$ are far from uniform. 

Both the JSD and cosine similarity treat all coordinates of $\simplex^{D}$ symmetrically. They are also only defined when  $p$ and $q$ have the same number of categories $D$. Alternatively, we may relax these constraints, requiring instead only a notion of pairwise similarity between coordinates in $p$ and $q$. This is formalized in optimal transport, which assigns costs for ``transporting mass'' between pairs of coordinates. Represent the costs of transporting mass between the $D$ coordinates of $p$ and the $D^\prime$ coordinates of $q$ by a matrix $C \in \reals_{+}^{D \times D^\prime}$. Then, the optimal transport between $p$ and $q$ is the coupling $\Pi$ minimizing
\begin{align*}
&\min_{\Pi \in \mathcal{U}\left(p, q\right)} \left<C,\Pi\right> \\
\mathcal{U}\left(p, q\right) := &\{\Pi\in \reals^{D \times D^\prime}_{+} : \Pi \*1_{D^\prime} = p \text{ and } \Pi^{T} \*1_{D} = q\},
\end{align*}
where $\left<A, B\right>$ is shorthand for the Frobenius inner product, $\text{tr}\left(A^T B\right)$. The smaller the transport cost $\left<C, \Pi\right>$, the more similar the distributions $p$ and $q$, with respect to the costs induced by $C$.

A useful analogy is due to Kantorovich \citep{peyre2019computational}. Imagine there are $D$ mines and $D^\prime$ factories. An amount $p_i$ of raw material is produced by mine $i$; on the other hand, factory $j$ requires $q_j$ total input. Suppose $C$ captures the transport costs between all pairs of mines $i$ and factories $j$. Then, the optimal transport plan $\Pi$ specifies how much material produced by mine $i$ should be shipped to factory $j$.

\section{Methods}
\label{sec:methods}

In this section, we set up the problem of topic alignment, provide associated algorithms, and discuss an R package implementation. Although more general treatments are possible, we focus on the case that the topics are derived from a sequence of models with increasing $K$. Alignment across a sequence of models supports multiscale analysis: topics from models with large $K$ distinguish between subtle variations in samples, and an alignment shows how these topics are related to overview topics derived at small $K$.
In Section \ref{sec:discussion}, we discuss how the methods proposed here could be generalized and applied for other purposes than multiscale analysis. For example, topic alignment could be used to compare topics identified in different environment (\textit{i.e.}, datasets) or across different modalities (\textit{i.e.}, different types of data have been collected on the same samples).

\subsection{Topic Alignment}
\label{sec:alignment}

Suppose we have estimated topics across an ensemble of LDA models $\mathcal{M}$. The topic alignment problem consists of constructing a weighted graph whose nodes are topics from across models and whose edge weights reflect the similarity between the topics. Formally, let $V$ be the set of topics across all models in $\mathcal{M}$. We suppose the investigator has specified pairs $e = \left(v, v^\prime\right) \in E$, where, $v, v^\prime \in V$ and $E$ is the set of edges in the topic alignment graph, of topics of interest to compare. Then, an alignment should provide weights $w: E \to \reals_{+} $ that are large when $v$ and $v^\prime$ have similar estimated parameters, and low otherwise.

The graph $\left(V, E, w\right)$ contains the result of the topic alignment. Let $k\left(v\right)$ denote the topic associated with node $v \in V$, and suppose it lies in model $m \in \mathcal{M}$. Write $\gamma\left(v\right) := \left(\gamma_{i k\left(v\right)}^m\right) \in \reals^N_{+}$ for the vector of mixed memberships associated with this topic. Similarly, set $\beta\left(v\right) := \beta_{k\left(v\right)}^m \in \simplex^{D}$.

\subsection{Algorithms}

\subsubsection{Weight estimation}
We propose two methods for estimating weights $w\left(e\right)$, one using sample composition ($\gamma_{i}$) and another using topic composition ($\beta_{k}$). We call the approaches \emph{ product alignment} and \emph{transport alignment}, respectively.

\begin{figure}
    \centering
    \includegraphics[width=\textwidth]{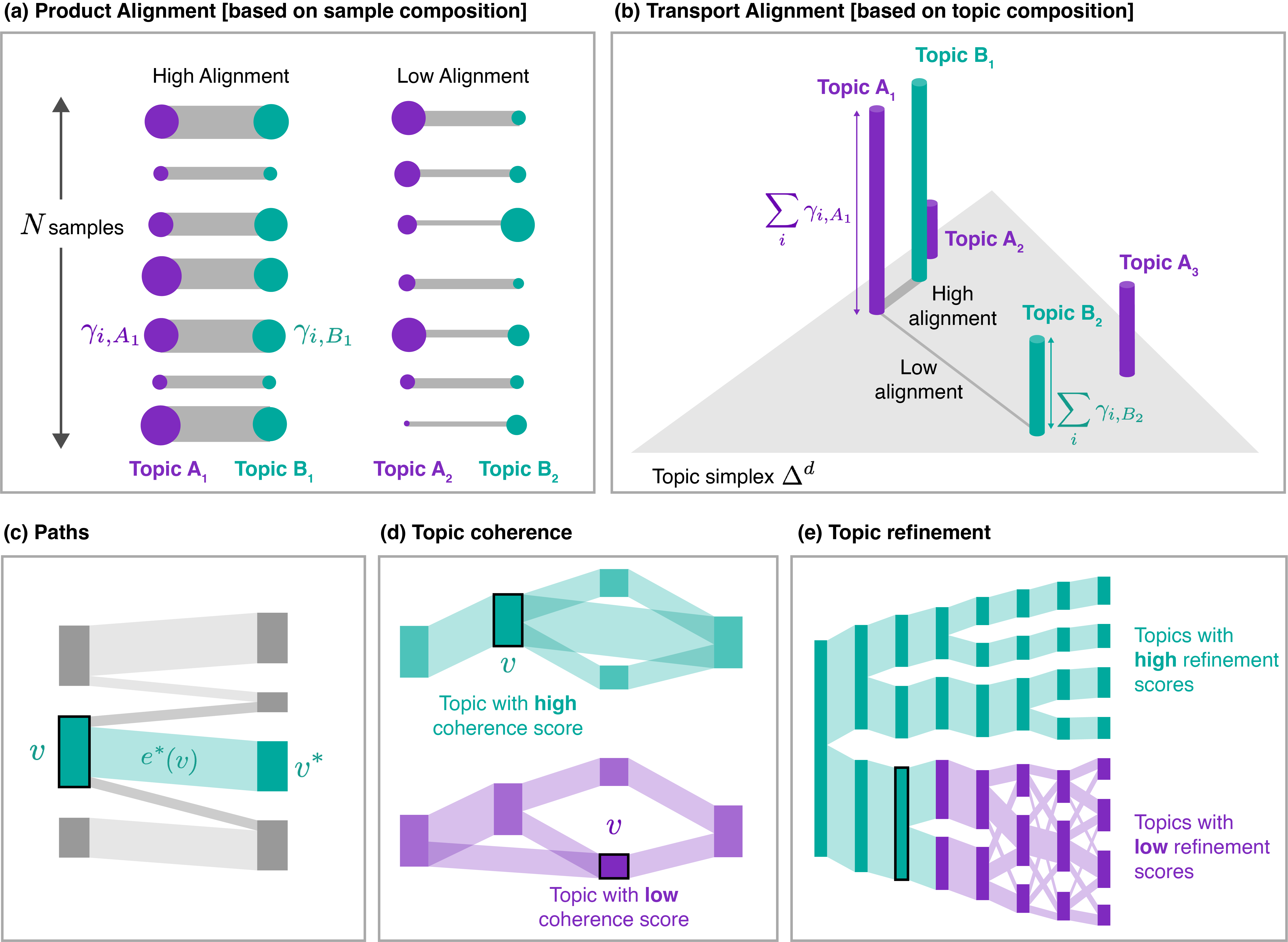}
    \caption{Top panels (a-b) illustrate topic alignment for product (a) and transport (b) alignments. The bottom panels (a-c) illustrate the diagnostic scores characterizing the alignment. (a) Each vertical column corresponds to a topic. Each circle encodes weights $\gamma_{iv}$ for a single sample $i$. The width of the links between circles encodes the product $\gamma_{iv}\gamma_{iv^\prime}$. Note that this product is large only if both $\gamma_{iv}$ and $\gamma_{iv^\prime}$ are large. The product alignment between two topics is high if the sum of products across all $N$ is large. 
    (b) Each vertical bar describes a single topic $v$. The heights of bars provide the weights $\sum_{i} \gamma_{iv}$ for each topic $v$; their locations encode $\beta_{v} \in \Delta^D$. Green and purple topics are estimated by LDA models with $K = 2$ and 3 topics, respectively. In transport alignment, the mass from the green bars is redistributed to the purple bars and alignment weights are derived from the associated optimal transport plan. (c) To assign a path to a topic $v$, the edges $e^\ast\left(v\right)$ from which topics $v$ derive most of their weight are identified. (d) A topic has a high coherence score if all normalized weights ($\win$ and $\wout$) between this topic and topics on the same path are large. (e) A topic has a high refinement score if the downstream alignment structure is ``tree-like'', i.e. if all descendant topics recognize $v$ as their main parent. Note that a topic $v$ (highlighted by a black outline here) may have a low coherence score but a high refinement score.}
    \label{fig:combined_alignment}
\end{figure}

In product alignment, we set $w\left(e\right) = \gamma\left(v\right)^T\gamma\left(v^\prime\right)$. Intuitively, if two topics have a similar pattern of $\gamma_{ik}$ across samples $i$, then they are given a high weight (Figure \ref{fig:combined_alignment}a). Further, topics that have small $\gamma_{ik}$ across all samples are given lower weight, regardless of their similarity.

In transport alignment, we compute $w\left(e\right)$ by solving a collection of optimal transport problems (Figure \ref{fig:combined_alignment}b). Consider two subsets $V_{p}, V_{q} \subset V$ with $V_{p} \cap V_{q} = \varnothing $; we take these two sets to be all topics $v$ from models $m$ and $m'$. 
Let $p = \left(\gamma\left(v\right)^T \*1_{N}\right)_{v \in V_{p}}$ and $q = \left(\gamma\left(v\right)^T \*1_{N}\right)_{v \in V_{q}}$. These summarize the ``mass'' of each topic across all samples, within each of the two sets. For example, these will both sum to $N$ if the $V_p$ and $V_q$ equal to the sets of topics from two models, since each $\gamma_i$ lies in the simplex. Define the cost of transporting mass from node $v$ to $v^\prime$ by $C\left(v, v^\prime\right) := JSD\left(\beta\left(v\right), \beta\left(v^\prime\right)\right)$. 
This ensures that weights are lower between topics with very different distributions, regardless of sample weights $\gamma_{ik}$.
Arrange these costs into a matrix $C$ of size $\absarg{V_p} \times \absarg{V_q}$. The weight matrix $W$ between pairs of topics in $V_{p}$ and $V_{q}$ is the $\reals^{\absarg{V_p} \times \absarg{V_q}}_{+}$ matrix formed by solving the transport problem
\begin{align*}
&\min_{W \in \mathcal{U}\left(p, q\right)} \left<C,W\right> \\
\mathcal{U}\left(p, q\right) := &\{W\in \reals^{\absarg{V_p} \times \absarg{V_q}}_{+} : W \*1_{\absarg{V_q}} = p \text{ and } W^{T} \*1_{\absarg{V_p}} = q\}.
\end{align*}

We note that in the case that $V_p$ and $V_q$ contain topics from models $m$ and $m + 1$, it is natural to construct a directed graph, with edges from topics in model $m$ to those in $m + 1$. In this case, we refer to the topic subsets as $V_{m}, V_{m + 1}$, respectively. For a directed graph, it is possible to normalize weights according to either the total inflow or outflow for each node. We will use these normalized weights in the computations of the topic orderings given in Supplementary Section 1 and in the computations of some of the diagnostic scores given in Section \ref{sec:diagnostics}. Specifically, we normalize weights for edges flowing out of $v$ according to $\wout\left(v, v^\prime\right) = \frac{w\left(v, v^\prime\right)}{\sum_{\tilde{v} : v \to \tilde{v}}w\left(v, \tilde{v}\right)}$. Similarly, normalization for edges flowing into $v$ is defined by $\win\left(v^\prime, v\right) = \frac{w\left(v^\prime, v\right)}{\sum_{\tilde{v} : \tilde{v} \to v} w\left(\tilde{v}, v\right)}$.

Figure \ref{fig:combined_true_lda} provides visualizations of product and transport alignments on simulated data. Note that topics are not returned by the LDA fit in a specific order. Consequently, topics connected by high weights across models may have different index $k$ within their respective model. For visualization purposes, we order topics within each model such that similar topics are close to each other. The ordering procedure is described in Supplementary Section 1.  

\subsubsection{Paths}
\label{subsec:paths}

Topic reordering places topics with high alignment weights next to one another, giving the appearance of chains of mutually similar topics.
To highlight this phenomenon, we partition the alignment graph into a collection of paths. The partition is grown iteratively, adding topics to existing subsets based on alignment weights.

Let $\text{Path}\left(v\right)$ be the path ID associated with topic $v$, and let $M$ be the model with the largest number of topics. For each topic $v \in V_M$, we initialize $\text{Path}\left(v\right) = k\left(v\right)$.
Suppose $\text{Path}\left(v\right)$ is known for all $v \in V_{m + 1}$. Then, the path membership $\text{Path}\left(v\right)$ of a node $v \in V_m$ is set to $\text{Path}\left(v^{\ast}\right)$, where
\begin{align*}
v^\ast &:= \arg\max_{v' \in V_{\left(m + 1\right):M}} (\wout\left(v, v^\prime\right) +  \win\left(v, v^\prime\right)),
\end{align*}
is the topic from one of the levels $m + 1, \dots, M$ that shares the highest total normalized weight with $v$.

\subsection{Diagnostics}
\label{sec:diagnostics}

We next propose three diagnostic measures that compactly describe the results of a topic alignment. These statistics reflect the added value of introducing each additional topic, the specificity of ancestor-descendant ties, and the coherence of topics across $K$. In addition to summarizing the alignment, these statistics can also serve to diagnose model mis-specification in the original fits.

\subsubsection{Number of paths}

Paths found by the iteration of Subsection \ref{subsec:paths} connect the most similar topics across resolutions. Spurious topics introduced at high resolution tend to be different from one another, limiting their ability to maintain a path. Instead, they connect to more stable paths. Consequently, counting the number of paths at a given resolution provides an indication of the number of true topics.  Formally, the number of paths for a model $m$ is the size of the set $\{\text{Path}\left(v\right) : v \in V_m\}$

In simulations below, we find that, when a topic model is appropriate, the true value $K$ is captured by a plateau in the number of paths (Figure \ref{fig:combined_true_lda}a). Hence, this metric can be used analogously to the identification of an ``elbow'' from a scree plot. Further, consistently slow growth in the number of paths identified may indicate departures from the assumed LDA model. Examples of both phenomena are provided in Section \ref{sec:simulations}. The number of paths is a property of a model within the alignment. In contrast, the scores introduced below focus on individual topics.

\subsubsection{Topic Coherence}

We call a topic \emph{coherent} if it is found in models fitted across a range of values of $K$. When coherent topics are recovered across multiple levels of an alignment, there is more evidence that the discovered structure is real, because it is not sensitive to the particular $K$ of the model used.

Topic coherence is defined in the context of paths. It measures the similarity between a given topic $v$ and the other topics on the same path $\mathcal{P}\left(v\right) = \{v^\prime: \text{Path}\left(v^\prime\right) = \text{Path}\left(v\right)\}$. It is defined as
\begin{align*}
c(v) = \frac{1}{|\mathcal{P}\left(v\right)|} \sum_{v' \in \mathcal{P}\left(v\right)} \min\left(\win\left(v, v'\right), \wout\left(v, v'\right) \right).
\end{align*}
Our simulations illustrate how this score can be used to identify ``good'' values of $K$ in LDA as well as detect departures from assumed LDA structure. Note that coherence focuses solely on the path containing a topic. We introduce another measure, topic refinement, to reflect the richer branching pattern downstream of a topic.

\subsubsection{Topic Refinement}\label{sec:refinement}

A topic identified at a small value of $K$ may have low coherence but still be a useful topic if it is the sole ancestor of topics in subsequent models. We expect true topics and compromises between true topics to have this property. We introduce the \emph{refinement} score to identify such topics.

Recall that for a node $v'$, $\win(v, v')$ measures the extent to which mass at $v'$ flows from parent node $v$. For each $v$, the refinement score is a weighted average of $\win(v, v')$ over all its children $v'$. More formally,  collect topics into levels $V_{1}, \dots, V_{M}$. We define the refinement score of node $v$ in level $m$ as
\begin{align}
\label{eq:refinement}
r\left(v\right) &= \frac{|V_m|}{M-m}\sum_{m'=m+1}^M \  \sum_{v_{m'}^\prime \in V_{m'}} \wout\left(v, v_{m'}^\prime\right)\win\left(v, v_{m'}^\prime\right).
\end{align}

To better understand this score, we can establish its properties in some simple cases (proofs given in the supplementary materials).
Continuing to assume that node $v$ is in level $m$, 
\begin{itemize}
    \item The refinement score is maximized ($r(v) = |V_m|$) if and only if $w(v, v'_{m'}) > 0$ implies $w(u, v'_{m'}) = 0$ for any $u \in V_m \setminus \{v\}$. This condition means that every descendant of $v$ has $v$ as its sole parent in level $m$. 
    \item The refinement score is minimized ($r(v) \to 0$) when all of the descendants of $v$ descend primarily from other nodes in level $m$ (i.e., the score is smallest for nodes that don't have any descendants that recognize them as parents at all). Indeed, suppose that for every $v' \in V_{m'}$, we have fixed weights $w(v, v')$. Then $r(v) \to 0$ when for each $w$ s.t. $w(v, v')> 0$,  $w(u, v') \to \infty$ for some $u \in V_m \setminus \{v\}$.
    \item The refinement score is defined such that $r(v) = 1$ if all the weights in the graph are equal, which indicates an absence of topic structure in the data.
\end{itemize}

\subsubsection{Comparing diagnostics}

The diagnostics measure different properties of an alignment. Both low coherence / high refinement and low refinement / high coherence combinations are possible, although in the examples below the diagnostics tend to track each other. We would expect the refinement score to be high but the coherence score to be low in the case that the alignment plot has a branching structure. On the other hand, the refinement score can be small for a topic with high coherence if that topic doesn't have many descendants. We discuss this further and provide examples in the Supplementary Section 4.

Overall, the coherence score describes how ``good'' or ``trustworthy'' a topic is; topics with high coherence scores appear consistently across levels. This is true even if the refinement score is low — in that case, the refinement score is likely to be low simply because the topic is present at low frequency. On the other hand, the combination of high refinement and low coherence score suggests that the topic is a mixture of several high-coherence topics. These topics can still be useful to the analyst, as they simply represent a coarser-grained summary of the data. 

\subsection{R package}
\label{subsec:package}

We have released an R package, \texttt{alto}, to support \emph{al}ignment of \emph{to}pics from LDA models. The package provides functions for
\begin{itemize}
\item Fitting a set of topic models.
\item Aligning topics across a collection of models, identifying paths, and computing coherence and refinement measures from the alignment.
\item Visualizing the resulting alignment object.
\end{itemize}
The design emphasizes the modularity of the alignment workflow, and separate functions are given for each of the steps above. To illustrate, we include an example use of the package on random multinomial data.
\begin{verbatim}
library(purrr)
library(alto)

# simulate data and fit models
x <- rmultinom(20, 5000, rep(0.1, 500))
lda_params <- setNames(map(1:10, ~ list(k = .)), 1:10)
lda_models <- run_lda_models(x, lda_params)

# perform alignment and plot
result <- align_topics(lda_models)
plot(result)
\end{verbatim}

Note that \texttt{result} is an S4 class (class \texttt{alignment}) with its own plot method. This class is associated with accessor functions for extracting the underlying model parameters (\texttt{models()}), alignment weights (\texttt{weights()}), and topic-level diagnostics (\texttt{topics()}).

In addition to the product and transport methods that are currently implemented, the package allows users to pass in arbitrary functions for computing weights between sets of topics. Further, in addition to computing alignments across a sequence of increasing $K$, the package implements topics comparison and weight construction over arbitrary topic graphs. Examples of these functions, as well as all data analysis and simulations described here, are available as package vignettes.
The package homepage is available at \url{lasy.github.io/alto/} and its source code can be found at \url{github.com/lasy/alto}.

\section{Simulations}
\label{sec:simulations}

In this section, we study the extent to which learned topic alignments and their associated diagnostics distinguish between types of variation that can arise in count data. We apply methods in a few controlled settings, verifying that derived interpretations are consistent with the known generative mechanism. We investigate alignment when simulating from true LDA models as well as under certain types of mis-specification. The latter cases inform the extent to which alignment can inform model assessment.

\subsection{Latent Dirichlet Allocation}
If the data were in fact simulated from an LDA model with $K$ topics, then what will the associated topic alignment and diagnostics look like?

We simulate $N = 250$ samples $x_i \in \naturals^{D}$ from an LDA model with $K = 5$ true topics and $D = 1000$. For mixed memberships, we draw $\gamma_{i} \sim \Dir\left(0.5 \cdot \*1_{K}\right)$, while topics are assumed sparser, with $\beta_{k} \sim \Dir\left(0.1 \cdot \*1_{D}\right)$. These parameters have been chosen to maintain simplicity while exhibiting both high-dimensionality in $x_i$ and sparse structure in $\gamma_i$ and $\beta_k$. At this scale, alignments can be made interactively: computation of product and transport alignments each takes 5 - 6 seconds on a laptop with a 3.1 GHz Intel Core i5 processor and 8GB memory (in contrast, to fit LDA models with $K \in \{2, \dots, 10\}$ requires 352 seconds).  We provide the true $\lambda_{\gamma}$ and $\lambda_{\beta}$ hyperparameters. In practice, these would be chosen
quantitatively according to marginal likelihood or qualitatively to enforce a desired level of sparsity. However, providing the true hyperparameters allows us to concentrate on the properties of alignment in an ideal case.

\begin{figure}
    \centering
    \includegraphics[width=0.95\textwidth]{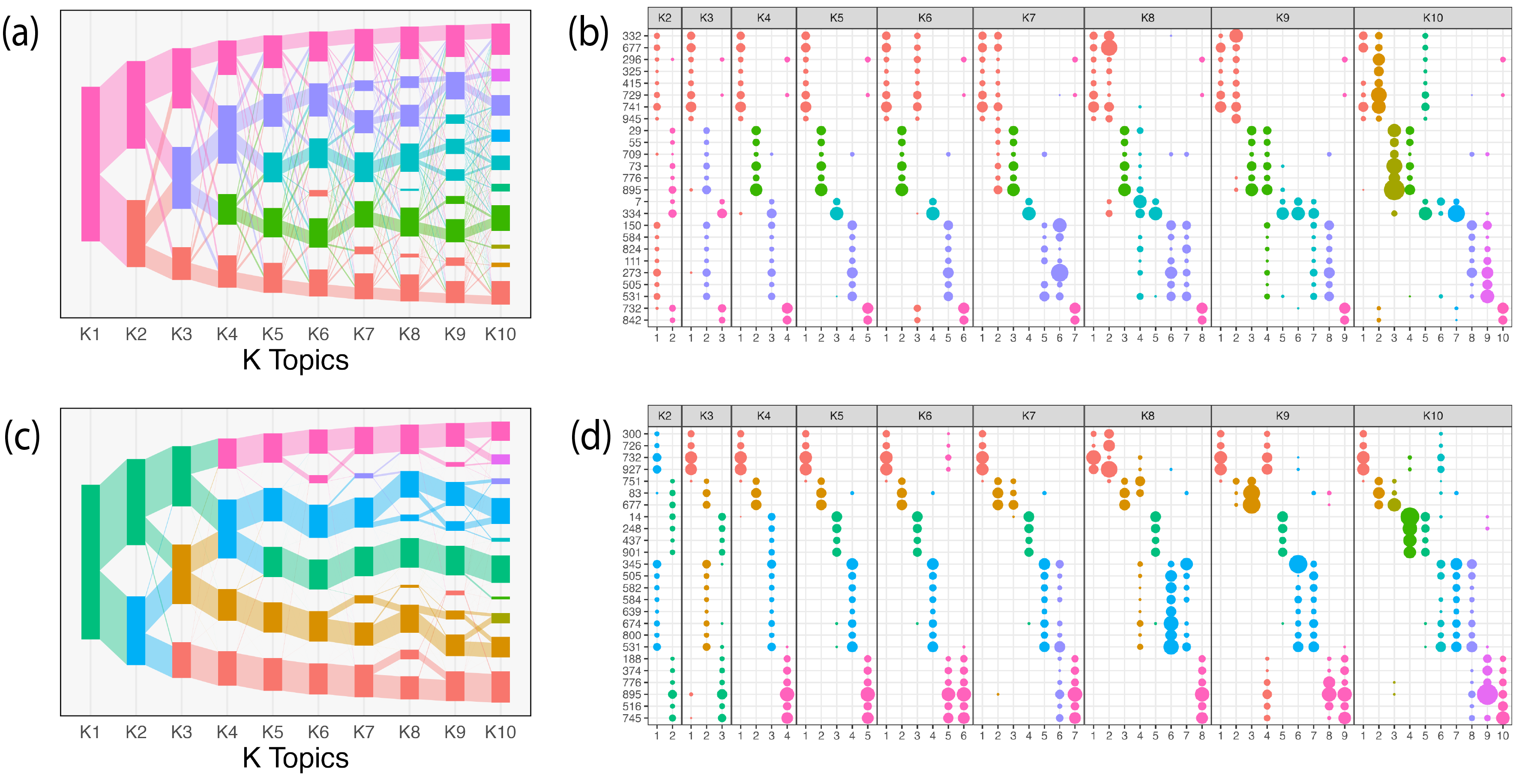}
    \caption{Alignments for data simulated from LDA with $K = 5$. Parts (a) and (c) are estimated using product and transport alignment, respectively. Rectangles correspond to topics, and their sizes give the mass $\sum_{i} \gamma_{ik}$. Vertical sections give fitted models. The width of links encodes the weights $w\left(e\right)$. Topics and edges are colored to show paths.
    Parts (b) and (d) give $\beta_{kd}$, colored in according to (a) and (c), respectively. Each column encodes a topic, each row is a dimension, and circle size is proportional to $\beta_{kd}$. Sets of topics from one model are grouped into panels. Circles with $\beta_{kd} < 0.001$ are omitted. Dimensions $d$ are sorted according to $\text{Distinctiveness}\left(d\right) := \min_{l \neq k} \beta_{kd} \log \frac{\beta_{kd}}{\beta_{ld}}+\beta_{ld}-\beta_{kd}$, as in \citep{dey2017visualizing}, but with $k, l$ varying over topics from multiple models. Only the 25 most distinctive dimensions are displayed.}
    \label{fig:combined_true_lda}
\end{figure}

With this setup, we simulate 200 datasets and fit models with $K \in \left\{2,\dots, 10\right\}$ topics. Each set of models is aligned using both the product and transport methods. The product and transport alignments from a randomly chosen replicate are shown in Figure \ref{fig:combined_true_lda}. 
The primary distinguishing feature between product and transport alignments is the sparsity in weights estimated using the transport approach. Both alignments provide hints that $K = 5$:
\begin{itemize}
\item The number of paths (i.e. number of distinct colors) remains 5 for $K > 5$.
\item For $K \leq 5$, most mass is conserved along a few major paths. For $K > 5$, this structure fragments and each topic tends to align with multiple descendant topics.
\end{itemize}

Supplemental Figures 4 - 9 provide ten additional replicates, along with ten replicates of a simulation from a null model in which the counts are drawn from independent multinomials whose means come from a $\Dir(\*1_{D})$ distribution. Each of the three diagnostics are shown for each of the simulated datasets. Clear differences are visible across all diagnostics for the data generated under the null \textit{vs.} topic model. In the topic model, the number of paths generally plateaus at 5. In the null model, the number of paths continues to increase as we add more topics. The coherence scores are all around 0 and the refinement scores are all around 1 in the null model. In the topic model, topics with low coherence and low refinement scores emerge for $K > 5$. These topics are likely spurious. The other topics (matching the true topics) have high coherence and refinement scores.

We next present a more systematic description of the diagnostics  across all 200 simulation replicates. Figure \ref{fig:gradient-combined}a counts the number of paths at each $K$.  Up to $K = 5$, and for most simulation runs, each new topic created a new path. For $K = 5, 6$, nearly all alignments estimated that 5 paths were present, though for larger $K$, additional topics were sometimes added to this subset. Transport alignment tended to more frequently overestimate the number of paths. For example, transport alignment occasionally found up to 8 paths when $K = 9$, while product alignment rarely estimated more than 5 topics.

Figures \ref{fig:gradient-combined}b-c show topic-wise coherence and refinement scores as a function of $K$ in the alignment. The lower envelope of the distributions for both coherence and refinement scores show an abrupt drop-off for $K > 5$ across both alignments, reflecting the low coherence and refinement of newly estimated topics with less similarity to the $K = 5$ true topics. For $K < 5$, refinement scores remain high as topics in these models are parents of true topics.

Overall, three practical rules of thumb are (1) a plateau in the number of paths indicates that the true number of topics has been reached, (2) a rapid drop-off in coherence or refinement scores indicates low-dimensional structure, and (3) topics with high coherence or refinement scores are more likely to reflect true topic structure.

To evaluate the extent to which these practical rules guide selection of $K$, we have repeated these simulations with datasets of increasing sample size. We observe that the probability of choosing the true $K$ increases with the sample size (Supplementary Figure 10).

These simulations are a sanity check — in the case that data are exactly generated by an LDA model with a known $K$, then alignment can help identify it. However, a number of methods are available for selecting $K$ when models are correctly specified, and real data are unlikely to so perfectly correspond to a proposed generative mechanism. In the spirit of ``all models are wrong, but some are useful,'' we consider, in the next two sections, scenarios where an LDA model is fit to data that are not simulated from the LDA mechanism, but where alignment can nonetheless inform an understanding of the essential latent structure.

\subsection{LDA with background variation}

To begin describing properties of alignment in this approximate regime, we simulate data from the case where sample compositions exhibit an extra level of heterogeneity not present in LDA. We suppose that most, but not all, variation in latent sample compositions lies on a $K$-dimensional subspace spanned by $K$ topics $B$. The closer the compositions lie to this subspace, the closer the LDA model is to being correct. Specifically, we simulate from
\begin{align*}
x_{i} \vert B, \gamma_{i}, \nu_i &\sim \Mult\left(n_{i}, \alpha B\gamma_{i} + \left(1 - \alpha\right)\nu_i\right) \\
\nu_{i} &\sim \Dir\left(\lambda_{\nu}\right) \\
\gamma_i &\sim \Dir\left(\lambda_{\gamma}\right) \\
\beta_{k} &\sim \Dir\left(\lambda_{\beta}\right).
\end{align*}
This generative mechanism is identical to that of LDA, except that instead of being centered around $B\gamma_{i}$, sample $i$ is centered around $\alpha B\gamma_i + \left(1 - \alpha\right)\nu_i$ for a $\nu_i\in \Delta^{D}$  drawn without reference to the $K$ topics in $B$. As before, we simulate with $N = 250, D = 1000, K = 5$. For each $\alpha \in \{0, 0.05, \dots, 1\}$, we generate 50 datasets and then fit and align topic models with $K \in \{1, \dots, 10\}$. 

Randomly chosen alignments for a range of $\alpha$ are given in Supplementary Figure 2. For large $\alpha$, most mass is concentrated in 5 core paths, and there is limited exchange from one topic to another. For small $\alpha$, mass is more evenly distributed across branches and a high-degree of exchange is present.

\begin{figure}
    \centering
    \includegraphics[width=0.8\textwidth]{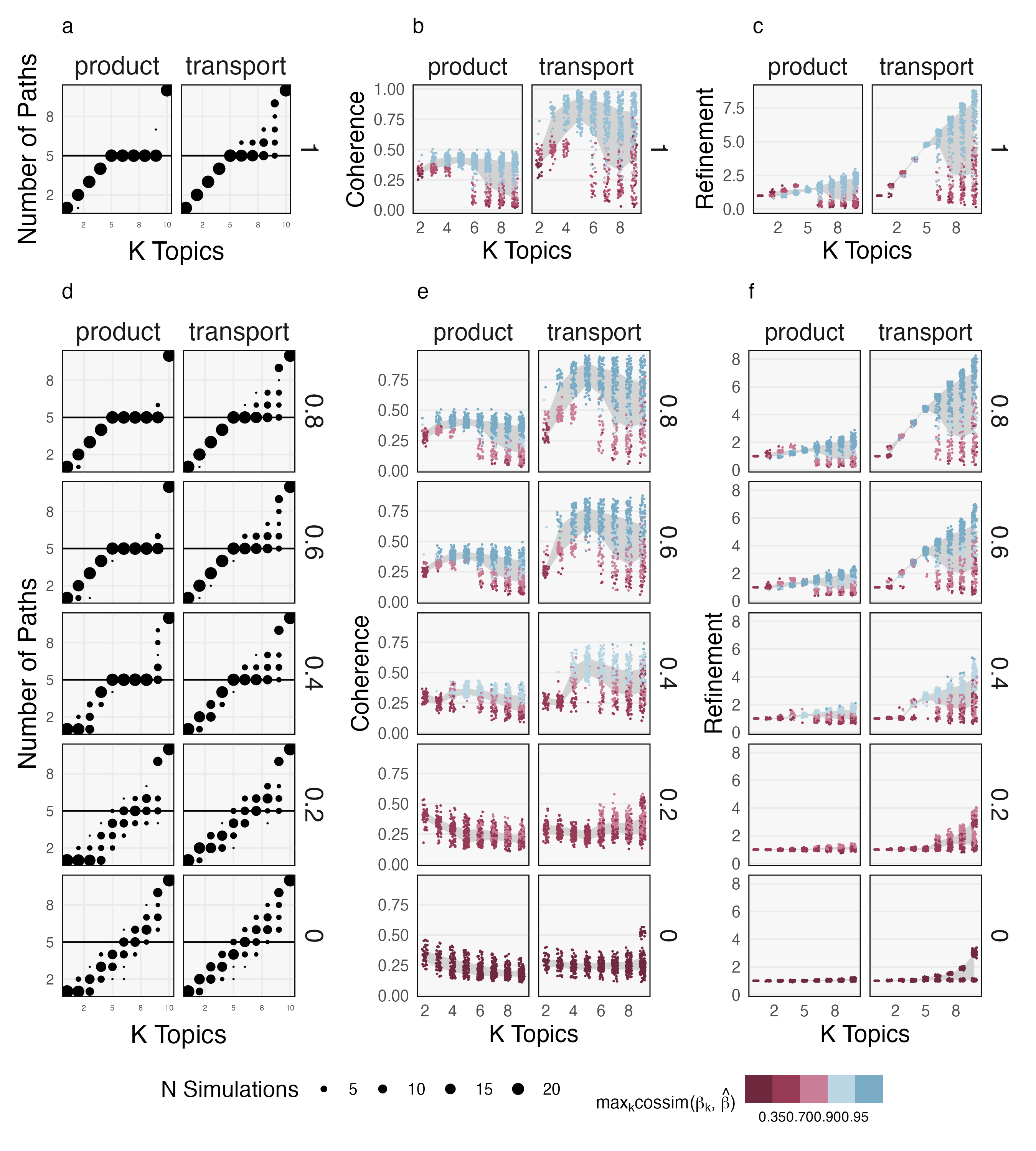}
    \caption{Diagnostic measures when background variation is and is not present. a) The number of estimated paths across simulations from an LDA model with $K = 5$. The circle size encodes number of replicates for which that number of paths was identified. The product method tends to be more conservative, and is less prone to overestimate the number of topics, compared to the transport method. 
    b) Coherence and c) refinement scores for topics fitted to data from an LDA model. Points represent estimated topics from across replicate. Color encodes similarity to a true underlying topic, which would be unknown in reality. d) The estimated number of paths varies as a function of background variation $\alpha$. The closer the data are to being drawnfrom an LDA model, the faster the initial increase in the number of estimated paths and the more definitive the plateau. 
    e) For small $\alpha$, coherence drops-off starting at $K = 1$, with no visible increases. For larger $\alpha$, a subset of topics has elevated coherence and the largest average topic coherence occurs at the true latent dimensionality. f) Refinement scores are higher and exhibit larger range when the LDA model is approximately correct. The range and trend refinement scores can be used to distinguish between datasets that have more or less unmodeled heterogeneity.}
    \label{fig:gradient-combined}
\end{figure}

The number of paths across $K$ for each $\alpha$ is shown Figure \ref{fig:gradient-combined}c. At $\alpha = 0$ (data simulated from random multinomials), there is no plateau in the number of paths. As $\alpha$ increases, a plateau at $K = 5$ emerges and becomes increasingly well-defined. 
The definition of paths appears effective at distinguishing low- from high-rank sample compositions — a gradual increase in the number of paths, without any visible plateau, would suggest that an LDA model is missing true sample-to-sample variation, even with large choices of $K$.

The distribution of coherence scores also shows differences depending on $\alpha$ (Figure \ref{fig:gradient-combined}e). For large $\alpha$ (\textit{i.e.}, generative mechanisms closer to LDA), the upper envelope of coherence scores rapidly increases up to $K = 5$. For $K > 5$, the lower envelope rapidly drops off while the upper envelope slightly decays. For small $\alpha$, there is no local maximum in the distribution of coherence scores and all topics have small scores.
This suggests that, when an underlying LDA model is more approximately correct, the associated alignment includes more coherent topics with a peak coherence around the true latent dimensionality.

Figure \ref{fig:gradient-combined}f shows the analogous display for refinement. In the small $\alpha$ case (no true topics), all topics have essentially the same refinement score for all $K$, and the score is as expected if there is no relationship among the topics. In contrast, in the more approximately low rank, large $\alpha$ case, a larger spread in scores is visible. In that case, topics with high refinement scores for $K \geq 5$ have high similarity with the true topics, and the $K = 5$ transition is marked by a drop-off in the lower envelope of refinement scores. Further, reading each panel from bottom to top (increasing topic structure), we find that the upper envelope of refinement scores noticeably increases.

Altogether, these diagnostics suggest that alignment can detect departures from the underlying topic model assumption that samples are concentrated on a $K$-dimensional topic simplex, across a range of candidate $K$. Paths with low coherence can be a warning flag. Further, low refinement scores and the absence of any plateau in the number of paths may suggest that observations exhibit higher sample-to-sample variation than an LDA model alone may capture.

Since it is possible to simulate new data from each fitted LDA model, these guidelines can be formalized into a graphical posterior predictive check. Posterior predictive samples can be drawn from the model with the largest $K$, which has the most flexibility. Aligning topic models fit to these data could provide a reference distribution for each of the diagnostics, and comparing the observed measures with this reference can give evidence for or against model fit.

\subsection{Strain switching}
\label{subsec:strain_switching}

Our final simulation studies whether alignment can detect mis-specifications in topic modeling due to highly correlated topics. Our setup is motivated by the strain switching phenomenon observed in some microbiota environments \citep{jeganathan2021statistical}. In this situation, there are strains that can be exchanged between what are otherwise similar communities. These strains can be thought of as being functionally equivalent, competing for a niche within an ecosystem. The consequence is that two nearly identical communities may be present in the ecosystem, but with systematic differences for some strains. From a topic modeling perspective, these communities have anti-correlated topic memberships -- only one of the competing strains can be present in a sample at a time.

The existence of these communities can be detected by comparing topics estimated at different scales. At coarse scale, two communities may be indistinguishable from one another, swamped by larger variations in species signatures across the ecosystem. At finer scale, the subsets of strains that distinguish them may become apparent after close inspection of the estimated topics. 

Our goal is to study the extent to which alignment can support this multiscale analysis. Our simulation mechanism first draws $\gamma_{i}$ and $\beta_{k}$ as in the simulations above. Instead of directly using $\beta_{k}$, however, perturbed versions $\tilde{\beta}_{k}^{r}$ are generated for $r \leq R_{k}$, a pre-specified number of perturbed replicates $R_{k}$. The perturbation mechanism is given in Supplementary Algorithm 1. The resulting $\tilde{\beta}_{k}^{r}$ and $\tilde{\beta}_{k}^{r^{\prime}}$ differ only on a subset of $S$ coordinates and can be viewed as functionally equivalent sub-communities. Given perturbed topics, sample $i$ is drawn by first randomly selecting one perturbed version from each of the $K$ topics,
\begin{align*}
\beta_{k}^{i} &\sim \Unif\left(\left\{\tilde{\beta}_{k}^{1}, \dots, \tilde{\beta}_{k}^{R_k}\right\}\right)
\end{align*}
binding the results into a $K$ column matrix $B_{i}$, and then drawing
\begin{align*}
x_{i} &\sim \Mult\left(n_{i}, B_{i}\gamma_{i}\right)
\end{align*}
as in standard LDA.

We set $K = 5$ and $\left(R_{1}, \dots, R_{5}\right) = \left(2, 2, 1, 1, 1\right)$. We draw $N = 250$ samples with dimension $D = 1000$. We use  $S = 230$; results with varying $S$ are given in the supplement. In the microbiota interpretation, our samples include counts of 1000 species each, and 5 underlying community types are present. Two versions of the first two types are present, differing on a subset of $S$ competing species.

\begin{figure}
    \centering
    \includegraphics[width=0.8\textwidth]{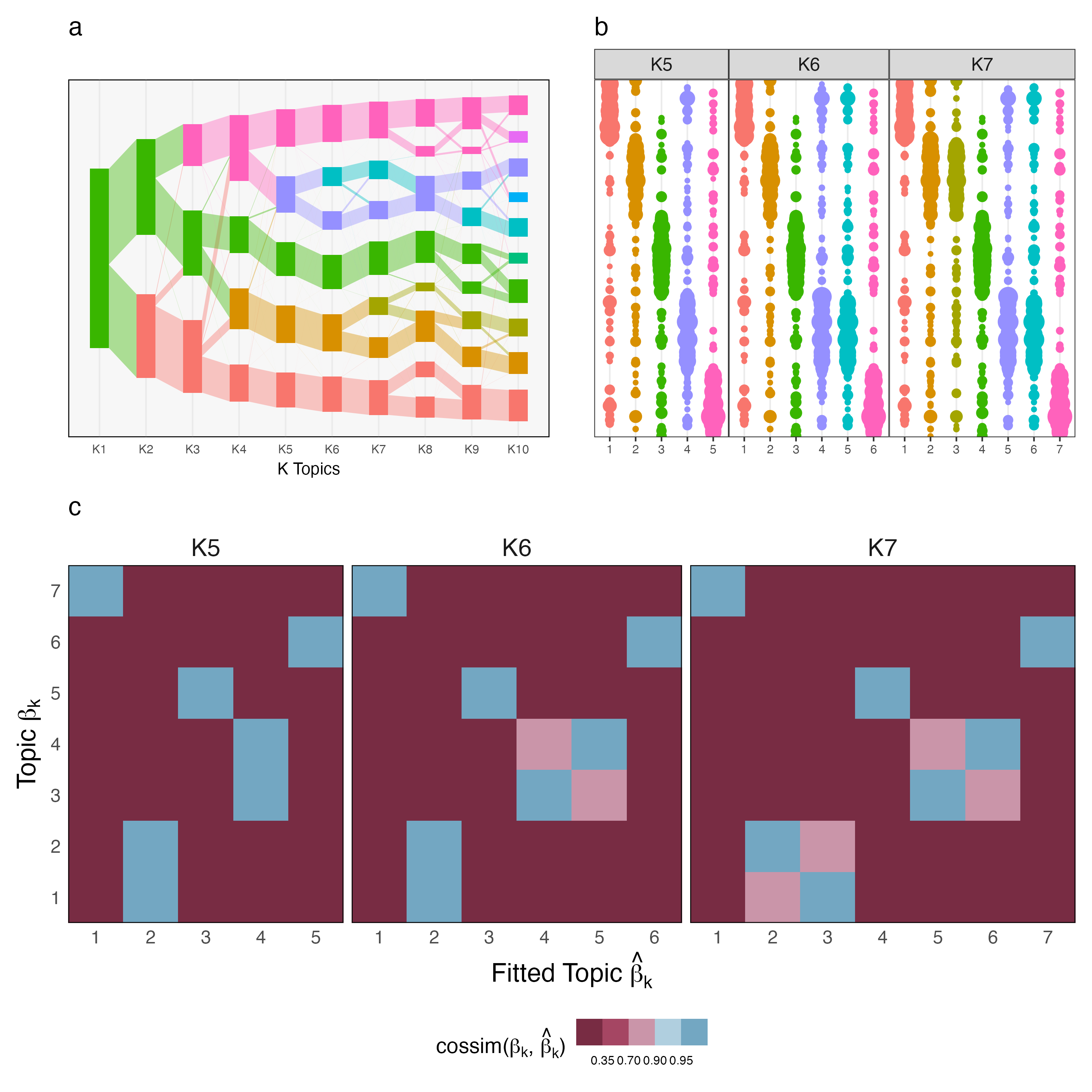}
    \caption{Results from the strain switching simulation. a) An alignment from one replicate with $S = 230$. Only the 200 most distinctive dimensions are displayed. The purple-dark blue and green-light blue pairs of branches correspond to two perturbed versions of the same underlying community, as suggested by b) the similar columns of $\beta_{kd}$ for $K = 6,7$. c) Cosine similarities between known and estimated topics across increasingly finer-scale models $m$. Rows 1-2 and 3-4 corresponding to perturbed versions of two underlying communities. For $K = 5$, the estimated topics do not distinguish between versions. At $K = 6$, rows 3 and 4 are slightly distinguished from one another, and at $K = 7$, both sets of perturbed topics are detected.}
    \label{fig:equivalence-combined}
\end{figure}

The resulting alignment is given in Figures \ref{fig:equivalence-combined}a-b. The learned topics for $K = 5$ to $7$ are given in the right panel.
We note that, at $K = 6$, the purple and blue topics have similar weights across many, but not all species. Likewise, at $K = 7$, the brown-orange and brown-green topic signatures are similar. 
The accompanying flow diagram shows that, in both cases, the pairs of similar topics had been merged when $K = 5$, suggesting that the model begins to detect perturbed versions of the same topics once $K$ is increased.

To attribute these differences to the known perturbation mechanism, we compute the cosine similarity between estimated and true topics. Figure \ref{fig:equivalence-combined}c shows the cosine similarity $\xi_{kk'}^m := \text{cossim}\left(\beta_{k}, \hat{\beta}_{k^\prime}^m\right)$ for models $m$ with 5 to 7 topics. Each row corresponds to a true topic; rows 1-2 and 3-4 are perturbed versions of two underlying sub-communities, respectively. For $K = 5$, the patterns of cosine similarities across rows 1-2 and 3-4 are similar, suggesting that the estimated topics are not sensitive to strain switching. However, for $K = 6$ and 7, new topics emerge that distinguish between the pairs of nearly equivalent sub-communities. The off-diagonal elements for the two squares indicates that the newly estimated topics remain similar to both versions of the underlying mechanism. However, since only a subset of species is perturbed in each version, some remaining similarity is to be expected.

\section{Data Analysis}
\label{sec:analysis}

\begin{figure}
    \centering
    \includegraphics[width=\textwidth]{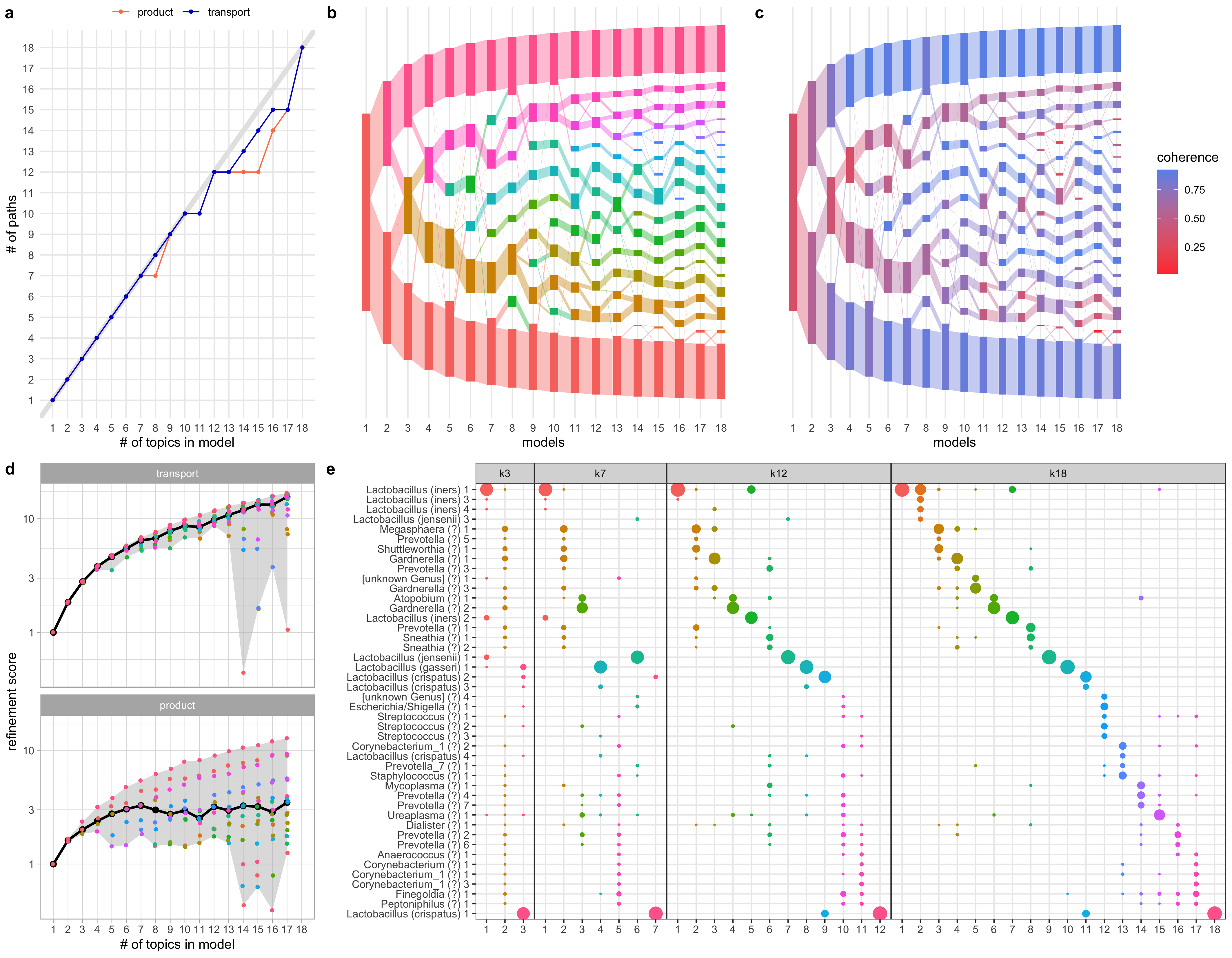}
    \caption{(a) Number of paths for each number of estimated topics in the LDA model. Number of paths identified by the transport method are shown in blue, those identified by the product method are shown in red. (b-c) Transport alignment of topics across $K$ where topics are colored by paths (b) or by their coherence score (b). (d) Small colored dots show the refinement score of each topic across $K$ for both alignment method. Colors match path colors in panel (b). The gray ribbon shows the envelope of the coherence scores (min to max), while the thick black lines shows the coherence scores median value. (e) Topic composition (dot size shows estimated $\beta$) for $K \in {3,7,12,18}$. Topics ($x$-axis) are colored by path (see panel b). Species ($y$-axis) are ordered by the topic with the highest $\beta$ for that species in model $K = 18$.}
    \label{fig:microbiota_figure}
\end{figure}

We applied topic alignment to vaginal microbiota composition data; the results are given in Figure \ref{fig:microbiota_figure}. The data are ASV counts from longitudinal samples collected throughout pregnancy in 135 individuals \citep{callahan_replication_2017}. In most individuals, the vaginal microbiota have low heterogeneity compared to other human microbiotas: one of four Lactobacillus species (\textit{crispatus}, \textit{iners}, \textit{gasserii} or \textit{jensenii}) completely dominates the flora. However, some individuals may present ``dysbiosis,'' defined by a high compositional diversity and the absence of Lactobacillus dominance. Topic analysis offers an opportunity to identify sub-communities that may co-exist within these diverse non-\textit{Lactobacillus} communities.

Applying topic alignment to these data, we observe that the number of paths (Figure \ref{fig:microbiota_figure}a) shows a small plateau around $K = 12$ with both methods (product and transport). As in the simulations, the number of paths are lower and the plateau is stronger when paths are identified using product rather than transport alignment. A small plateau is likely indicative that the data generation process does not strictly follow the LDA model assumption. However, most of the identified topics around $K=12$ are coherent across $K$ (Figure \ref{fig:microbiota_figure}b-c). The distribution of refinement scores (Figure \ref{fig:microbiota_figure}d) shows, for both alignment methods, the emergence of low refinement score topics from $K = 14$. This supports the idea that a higher number of topics is likely over-fitting the data. Further, the median refinement score is highest for $K = 7$ when using the product method. This suggests that topics identified at that resolution are a mixture of true, higher resolution topics. 

In summary, a biologist may interpret this analysis by stating that topic models with $K = 12$ provides the best summary of the sub-communities found in the vaginal microbiota. Among those 12 sub-communities, two (topic 11 and 12, with low coherence scores) might be ``spurious'' in the sense that they may not represent well-defined sub-communities but instead capture a set of bacteria that may be sample-specific (background noise). The analyst could also choose to model their data at a coarser resolution by setting $K = 7$. At that resolution, they would identify four coherent \textit{Lactobacillus}-dominated topics and three non-\textit{Lactobacillus} dominated topics. Among these three topics, one topic has a high coherence score and is composed of  specific species of Gardnerella and Atopobium. The other two topics, with lower coherence scores but high refinement scores, identify two distinct mixtures of sub-communities which are revealed at higher resolution.

These results are useful from a biological perspective because they provide a more detailed, and yet still succinct, description of the vaginal microbiota structure. Historically, vaginal microbiota data have been clustered into five community state types (CST), four of them corresponding to one of the four most prevalent \textit{Lactobacillus} species, and the fifth one being ``everything else.'' Our analysis is consistent with this interpretation, but provides a more precise view into the structure of the fifth state.

\section{Comparison with alternatives}
\label{sec:alternatives}

In this section, we present and discuss analyses comparing topic alignment with alternative methods (perplexity) or models (hierarchical LDA).

\subsection{Perplexity}

Here, we evaluate train and test perplexity for each fitted model across simulations. Perplexity is a measure of the probability of test samples under a fitted model (see supplementary equation 4.1). For each simulation setup, we compute perplexity both on the data used to train the model and an independent sample with the same topics $\beta_{k}$. Supplementary Figure 11 shows that, when the data are generated via LDA, an ``elbow'' in train and test perplexities highlights the correct choice of $K = 5$. In the case of data generated with background noise (Supplementary Figure 12), a subtle drop-off around $K = 5$ is visible at small $\alpha$ and grows more apparent and concentrated around $K = 5$ as the noise decreases. For strain switching (Supplementary Figure 13), an elbow at $K = 5$ is visible, but even for large $S$, no indication of switching emerges. In each case, test perplexity never increases after the true $K = 5$, but the location of the ``elbow'' nonetheless suggests the correct $K$, in most cases.

While perplexity can be used to inform the selection of the number of topics, alignment can provide relevant, complementary information. For example, perplexity is defined on subsets of samples, and so, unlike coherence or refinement, it cannot be used to evaluate the quality of individual topics. Further, in the case that the optimal perplexity appears at a large $K$, it can still be worthwhile to use topics at a smaller $K$ to guide interpretation of aligned topics at the optimal, larger $K$. Perplexity alone does not support such a details-on-demand analysis. Finally, though subtle differences in perplexity curves for true LDA vs. mis-specified models are apparent (e.g., subtle decreases in strain-switching perplexity after $K = 5$), variations 
across types of mis-specification are more clearly evident through model alignment.

\subsection{Hierarchical LDA (hLDA)}

In this section, we contrast the proposed topic alignment with hierarchichal LDA \citep{blei2003hierarchical}. 
While topic alignment visualizations are similar to visualizations of hierarchical structures, it is important to note that topic alignment is not a hierarchical method. Topic alignment relies on different assumptions and fulfills a different purpose than hierarchical topic models (hLDA). Applying these methods on the same datasets leads to different results, interpretations, and conclusions.  

First, hLDA assumes that topics follow a strict tree-like hierarchical structure: child-topics have only one parent. In contrast, the alignment structure is not tree-like and topics at higher resolution may be connected to several topics at lower resolution. Second, in the hLDA framework, samples belong to a single path in the hierarchy; they can only be composed of topics that are part of the same branch. In contrast, topic alignment only describes relationships between topics at different resolutions. Within a resolution, samples are described as mixture of topics. Third, because hLDA is a more complex model, it has two additional hyperparameters (the depth, and the concentration parameter for introducing new topics) compared to LDA. Consequently, deploying hLDA on datasets requires additional effort to identify optimal values for these parameters. Finally, in hLDA, we can interpret child-topics as sub-topics, and the hierarchical structure as a topic taxonomy. For example, in the context of analysing a magazines corpus, \textit{football}, \textit{tennis}, and \textit{climbing} could be sub-topics of a \textit{sport} topic. Specific terms characterize these sub-topics (e.g., ``harness'' for \textit{climbing}, or ``racket'' for \textit{tennis}), while the \textit{sport} parent topic might be characterized by terms such as ``competition,'' ``training,'' or ``fitness'' which we expect to find in documents related to either \textit{football}, \textit{climbing}, or \textit{tennis}. Topic alignment may also lead to a similar interpretation of topic relationships, but exclusively for topics with high refinement scores. 

Topic alignment is a post-estimation method aimed to guide scientists in their exploratory analyses when modeling their data with topic models. There are no assumptions regarding the relationships between topics at different resolutions. These relationships and the diagnostic scores provide information to users for interpreting their data. This is especially useful if the data generation process does not strictly follow the LDA assumptions and when perplexity curves do not show a clear elbow.

Importantly, for microbiota structure analyses, the hLDA assumptions are not in agreement with observed data and current understanding of the microbial biology. Even if bacterial sub-communities were organized hierarchically (e.g., because of strain switching) we would still expect sub-communities from different branches of the hierarchy to co-exist within a given ecosystem (i.e., within a sample). This results in more complex interpretations; Supplementary Figure 11 demonstrates how hLDA introduces a degree of redundancy to account for mixtures across branches (see Supplementary Figure 11). 

Finally, current implementations of hLDA \citep{tomotopy} are not well suited for analyses of microbiota composition for two practical reasons. First, they require samples to be provided in a corpus format, as opposed to a matrix of counts. Given current library depths, transforming ASV counts into text files leads to large files (7+ GB). Second, the time required to fit a single hLDA model is larger than to fit LDA models at multiple resolutions and aligning the topics. For example, fitting hLDA on a subset of the vaginal microbiota data takes just under a minute.
In comparison, it takes approximately 20-25 seconds to fit 15 LDA models and 
perform the topic alignment on the same dataset.

\section{Discussion}
\label{sec:discussion}

We have introduced techniques for aligning topics across an ensemble of topic models. The resulting estimates provide a multiscale view of count data, showing how topics from large and small $K$ models compare and contrast with one another. We framed the alignment problem as the construction of the appropriate weighted graph whose nodes represent topics and whose edges encode topic similarity. We provided algorithms for estimating weights based on either the inner product or the optimal transport between fitted model parameters. Based on these alignment weights, we proposed diagnostics describing (1) the extent to which any given topic persists across a range of $K$ (coherence score) and (2) the definitiveness with which finely-resolved topics emerge from coarser ones (refinement score). We studied the properties of the proposed methods through a series of simulations, emphasizing the potential for alignment to detect biologically plausible departures from the LDA generative mechanism. We also applied the overall workflow to a vaginal microbiota dataset and recovered both known, high-level CSTs, and novel finer-grained sub-community structure.

We note several limitations and opportunities for future study. We have not provided any theoretical guarantees about the estimated alignment weights or diagnostics. In order to make our approach applicable to the ensembles of fitted LDA models that are most frequent in practice, we have deliberately avoided proposing an overarching multiscale model. Requiring a new model would increase the burden for adoption -- it is easier to compute post-estimation statistics within a familiar workflow. Nonetheless, though beyond our scope, it would be worthwhile to understand the behavior of alignment weights or diagnostics in such a multiscale setting where model parameters are assumed to be drawn from a plausible distribution.

Further, we have not incorporated any interactive visualization principles to streamline the analysis of the final alignment data structure. The static views provided by our package describe a single aspect of alignment at a time, showing the alignment weights, the estimated model parameters, and diagnostics in isolation from one another. It would be useful to link these views interactively. For example, the ``top'' species associated along each branch could be highlighted interactively, or the species whose distributions change the most from one topic to the next. Also absent from our views are any visualizations of how individual samples relate to the topic alignment overall.

Finally, we note that, though we have focused on the case of increasing $K$, the principle of computing summaries that characterize an ensemble of models is more generally applicable. For example, the choice of hyperparameters $\lambda_{\gamma}, \lambda_{\beta}$ controls the sparsity of the posterior mixed membership and topic estimates. A view of which estimates are most strongly influenced by these hyperparameters would be informative. Further, in the data integration context, it may be simpler to relate separate models fit across data modalities rather than to construct a new global model for each new combination of component modalities. Similarly, for datasets collected across multiple sites or environments, alignment may provide a compromise between fitting a separate model per site, which fails to pool any shared information, and implementing a full hierarchical model, which can be a labor-intensive exercise. In these cases, the sets $V_{p}$ and $V_{q}$ for alignment contain not just topics from adjacent models, but topics from across a larger ensemble.

As the types of data incorporated in biostatistical studies grow in number and complexity, flexible techniques for dimensionality reduction and visualization will continue to be an important component of the data analysis workflow. Exploratory analysis can guide the critical examination of complex problems, and topic alignment is a simple but useful addition to the toolbox available for count data.

\section{Software}
\label{sec5}

The R package \texttt{alto} is available at \url{lasy.github.io/alto}. Simulations and data analysis can be reproduced through package vignettes. Scripts for reproducing simulations in a high-performance computing environment are available at \url{github.com/krisrs1128/topic_align}.

\section{Supplementary Material}
\label{sec6}

Supplementary figures, algorithms, and proofs are available online at \url{http://biostatistics.oxfordjournals.org}.

\section*{Acknowledgments}
The authors thank Prof. Susan Holmes and Prof. Karl Rohe for fruitful discussions and constructive feedback on the manuscript. 
{\it Conflict of Interest}: None declared.

\section*{Funding}
This work was supported by the Bill and Melinda Gates Foundation grant OPP1189205-2019 (L.S.). 

\bibliographystyle{biorefs}
\bibliography{refs}

\setcounter{section}{0}
\section*{Supplementary Materials}

These supplemental materials provide further theoretical and experimental results that do not appear in the main paper. In more detail, these supplemental sections describe,
\begin{itemize}
\item Section 1: A table of notation.
\item Additional conceptual discussion of alignment and diagnostic measures,
\begin{itemize}
\item Section 2: A description of the topic reordering strategy used to ensure that alignment visualizations do not become tangled as models increase in resolution.
\item Section 3: Properties of the refinement score. Proves the maximization and minimization results given in the main manuscript. Also derives refinement scores in the case that weights are equal.
\item Section 4: Contrasting diagnostics. Provides a simple example where the refinement score is large, but not the coherence score, and vice versa.
\item Section 5: Comparisons of alignment diagrams from data drawn from true LDA and null multinomial generative mechanisms. Shades results in according to either path ID or the proposed diagnostic measures.
\end{itemize}
\item Further simulation results and commentary,
\begin{itemize}
\item Section 6: Provides alignment visualizations corresponding to the background noise simulation in the main text.
\item Section 7: Visualizes the convergence of diagnostic measures as the number of samples increases. Suggests a form of consistency, albeit in a limited case.
\item Section 8: Discusses strain switching properties across number of switched species S and all simulation replicates. Also provides specific algorithm for strain switching.
\end{itemize}
\item Discussion of related methods, supporting Section 6 of the main text.
\begin{itemize}
\item Section 9: Perplexity measures across simulation experiments. Provides a complementary approach to model selection.
\item Section 10: Application of hierarchical LDA to the vaginal microbiota dataset. Discusses interpretation of fitted parameters and contrasts this with our proposed alignment.
\end{itemize}
\end{itemize}

\section{Notation}

\begin{longtable}{rp{12cm}}
\textbf{Notation} & \textbf{Interpretation} \\ \hline
$N$ & The total number of samples. \\  
$D$ & The dimensionality of each sample. \\  
$x_i$ & The vector of counts for sample $i$. $x_i \in \naturals^{D} $\\  
$n_i$ & The total count of sample $i$. That is, $n_i = \sum_{d = 1}^{D} x_{id}$. \\  
$\Delta^{K}$ & The $K$-dimensional simplex. \\  
$\gamma_i$ & The topic memberships for sample $i$. $\gamma_i \in \Delta^{K}$\\  
$K$ & The number of topics. \\  
$\*1_{K}$ & A vector of $K$ ones. \\  
$\beta_{k}$ & The composition of topic $k$. $\beta_{k} \in \Delta^{D}$\\  
$B$ & A $D \times K$ matrix where the $k^{\text{th}}$ column is $\beta_{k}$ (composition of topic $k$). \\  
$\lambda_{\gamma}, \lambda_{\beta}$ & Hyperparameters of the Dirichlet distributions for $\gamma_{i}$ and and $\beta_{k}$, respectively. \\  
$p, q$ & In general, two distributions. In all examples, these correspond to the locations and weights associated with topics $v$ in the graph defining an alignment. \\  
$C$ & A matrix of transport costs from $D$ coordinates of $p$ to $D^\prime$ coordinates of $q$. In examples, C holds the JSDs between topics across two models. $C \in \reals_{+}^{D \times D^\prime} $ \\  
$\Pi$ & The optimal transport map between two distributions. $\Pi \in \reals_{+}^{D \times D^\prime}$\\  
$m \in \mathcal{M}$ & A single model ($m$) within the larger ensemble of all models ($\mathcal{M}$) \\  
$V, E$ & The vertices and edges representing topics across all models and the potential alignments between them. \\  
  $v$ & A specific topic within $V$. \\
 $e(v, v')$ & A pair of topics. $e \in E$ \\
$w$ & $: E \to \reals^{+}$ The alignment weights associated with pairs of models on $E$. \\ 
$w(e)$ & The alignment weight for the pair $e$. \\
$W$ & The matrix of weights $w\left(e\right)$ for all edges with a specified subset. \\  
$\win, \wout$ & Normalized alignment weights, when weights are associated with directed edges. \\  
$k\left(v\right)$ & The index of topic $v$ within a subset of $V_m$ of topics derived by model $m$. \\  
$\gamma\left(v\right)$ & $ \in \reals_{+}^{N}$ The vector of memberships $\gamma_{ik\left(v\right)}$ for topic $v$ across all $N$ samples $i$. \\  
$V_{p}, V_{q}$ & Two subsets of topics. When these are written as $V_{m}$ and $V_{m+ 1}$, these are subsets from two models with $m$ and $m + 1$ topics, respectively. \\  
$\psi_{m}$ & The permutation used to reorder topics for model $m$. \\  
$\Psi_{m}$ & The set of potential permutations for reordering topics in model $m$. \\  
$\text{Path}\left(v\right)$ & The (scalar) path identity associated with topic $v$. \\  
$\mathcal{P}\left(v\right)$ & The subset of vertices with the same path identity as topic $v$. \\  
$c\left(v\right)$ & The coherence score associated with topic $v$. \\  
$r\left(v\right)$ & The refinement score associated with topic $v$. \\  
$\nu_i$ & The sample-specific background noise multinomial parameter in the background noise simulation. \\  
$\alpha$ & The extent of LDA structure in the background noise simulation. $\alpha = 1$ gives a true LDA model, while $\alpha = 0$ corresponds to the multinomial null model. \\  
\hline
\caption{Glossary of notation used in this paper.}
\label{tab:notation}
\end{longtable}

\section{Topic ordering}

\begin{figure}[h]
    \centering
    \includegraphics[width=0.5\textwidth]{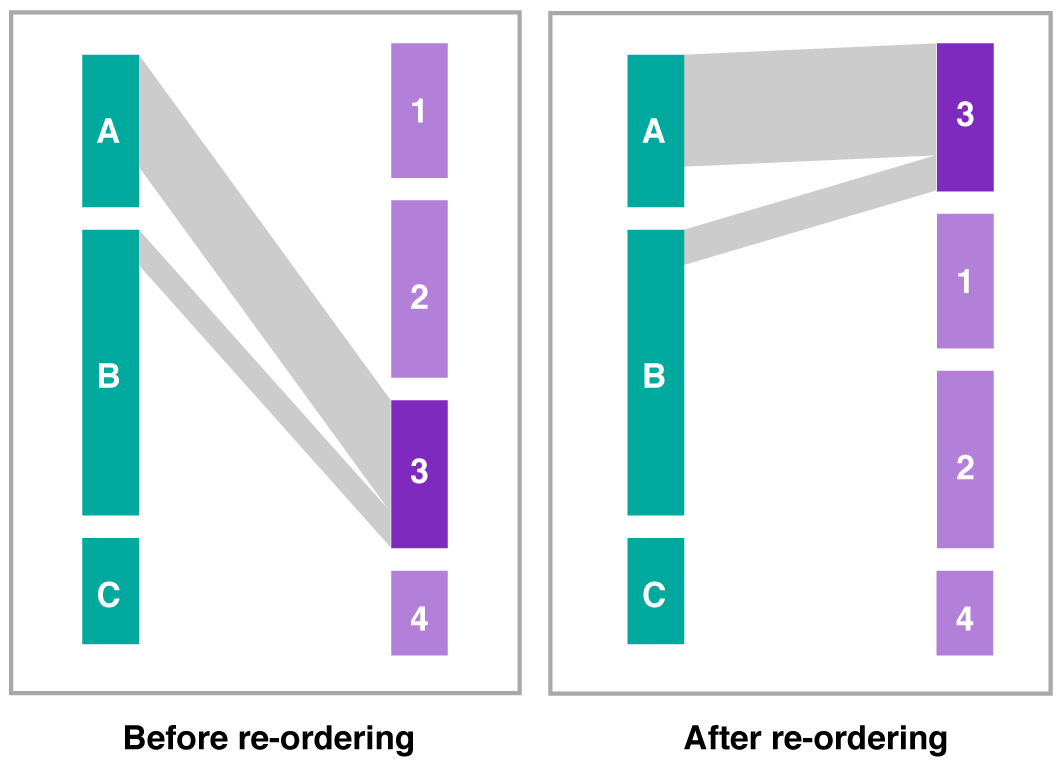}
    \caption{Given the high alignment weights with topic A, topic 3 index is permuted such that this topic become the first one of its model.}
    \label{fig:reordering}
\end{figure}

Topics are not returned by the LDA fit in a specific order. Consequently, topics connected by high weights across models may have different index $k$ within their respective model. For visualization purposes, it is useful to order topics within each model such that similar topics are close to each other (Supplementary Figure \ref{fig:reordering}). The ordering procedure seeks the optimal permutation of topic indices $\psi_{1:M}^\ast$ such that the distance between strongly connected consecutive topics is minimized:
\begin{align*}
    \arg \min_{\psi_{1:M} \in \Psi_{1:M}} \sum_{m = 1} ^{M - 1} \sum_{e \in E_{m, m + 1}} \absarg{\psi_{m}\left[k\left(v\right)\right] - \psi_{m + 1}\left[k\left(v'\right)\right]} w\left(e\right),
\end{align*}
where the optimization is taken over the set of possible topic permutations $\Psi_{m}$ of topic labels in each model $m$ and $E_{m, m+1}$ is the set of edges between topics in models $m$ and $m + 1$. Finally, the reordered topic label for node $v$ at level $m$ is given by $k\left(v\right) \leftarrow \psi_{m}^\ast\left(k\left(v\right)\right)$. For example, in Supplementary Figure \ref{fig:reordering},
suppose that the topics $v \in V_{m}$ for the purple topic have values of $k\left(v\right)$ of $1, \dots, 4$, arranged from top to bottom. Then, the associated permutation $\psi\left(v\right)$ is $\left(3, 1, 2, 4\right)$, also proceeding from top to bottom.

Instead of searching over all possible permutations, we approximate the optimal solution across a sequence of models $M$ by applying a forward and a backward pass, both of which rank the centers of gravity of a topic based on the weights connecting it to topics from the previous (forward pass) or next (backward pass) model. We find that additional forward and backward passes have little impact on the rankings. 
Specifically, the set of topic indices is updated using Algorithm \ref{alg:reorder}.

\begin{algorithm}[H]
\For{m = 2:M}{
$k^\prime\left(v_m\right) := \text{rank} \left(\sum_{v_{m-1} \in V_{m-1}} k\left(v_{m-1}\right) \win\left(v_{m-1}, v_{m}\right)\right),  \forall v_m \in V_m$
}
\For{m = M:2}{
$k^\prime\left(v_{m-1}\right) := \text{rank} \left(\sum_{v_{m} \in V_{m}} k\left(v_{m}\right) \wout\left(v_{m-1}, v_{m}\right) \right),  \forall v_{m-1} \in V_{m-1}$
}
\label{alg:reorder}
\caption{Forward and backward pass for the topic ordering algorithm. In the forward pass, topics are indexed so that they are close to the source topics from which they draw the most weight, while in the backward pass, they are placed near their high weight descendants.}
\end{algorithm}

\section{Properties of the refinement score}

Our definition of the refinement score is
\begin{align*}
  r(v) &:= \frac{|V_l|}{L - l} \sum_{l' = l + 1}^L \sum_{v'_{l'} \in V_{l'}} \wout(v, v'_{l'})\win(v, v'_{l'})\\
  &= \frac{|V_l|}{L-l} \sum_{l' = l+1}^L \sum_{v'_{l'} \in V_{l'}} \frac{w(v, v'_{l'})^2}{(\sum_{u \in V_l} w(u, v'_{l'}))(\sum_{w \in V_{l'}} w(v, w))}
\end{align*}
where $V_{l}$ is the set of all nodes in level $l$.

Here we give proofs for the assertions about the refinement score given in the main text.

\subsection*{Maximizing $r(v)$}
Suppose $v$ is in level $l$, and suppose further that $w(v, v'_{l'}) > 0$ implies $w(u, v'_{l'}) = 0$ for any $u \in V_l \setminus \{v\}$.
This means that every node in level $l'$ has only one parent in level $l$.
In that case, $\sum_{u \in V_l} w(u, v'_{l'}) = w(v, v'_{l'})$, and we can write the inner sum in the definition of $r(v)$ as
\begin{align*}
  \sum_{v'_{l'} \in V_{l'}} \frac{w(v, v'_{l'})^2}{(\sum_{u \in V_l} w(u, v'_{l'}))(\sum_{w \in V_{l'}} w(v, w))} &= \frac{1}{\sum_{w \in V_{l'}}w(v,w)} \sum_{v'_{l'} \in V_{l'}} \frac{w(v, v'_{l'})^2}{\sum_{u \in V_l} w(u, v'_{l'})} \\
  &=  \frac{1}{\sum_{w \in V_{l'}}w(v,w)} \sum_{v'_{l'} \in V_{l'}} w(v, v'_{l'}) = 1
\end{align*}
Then the overall value for the refinement score is
\begin{align*}
r(v) &= \frac{|V_l|}{L-l} \sum_{l' = l + 1}^L\sum_{v'_{l'} \in V_{l'}} \frac{w(v, v'_{l'})^2}{(\sum_{u \in V_l} w(u, v'_{l'}))(\sum_{w \in V_{l'}} w(v, w))}\\
&= \frac{|V_l|}{L-l}\sum_{l'=l+1}^L 1 = 1
\end{align*}

This is the largest $r(v)$ can be, as can be seen by noting that if $w(u,v'_{l'}) > 0$ for some $u \in V_l \setminus \{v\}$, we will have
\begin{align*}
  \sum_{v'_{l'} \in V_{l'}} \frac{w(v, v'_{l'})^2}{(\sum_{u \in V_l} w(u, v'_{l'}))(\sum_{w \in V_{l'}} w(v, w))} &= \frac{1}{\sum_{w \in V_{l'}}w(v,w)} \sum_{v'_{l'} \in V_{l'}} \frac{w(v, v'_{l'})^2}{\sum_{u \in V_l} w(u, v'_{l'})} \\
  &< \frac{1}{\sum_{w \in V_{l'}}w(v,w)} \sum_{v'_{l'} \in V_{l'}} \frac{w(v, v'_{l'})^2}{w(v, v'_{l'})} \\
  &=  \frac{1}{\sum_{w \in V_{l'}}w(v,w)} \sum_{v'_{l'} \in V_{l'}} w(v, v'_{l'}) = 1
\end{align*}

Therefore, a node will have a refinement score of 1 if and only if every node in level $l' > l$ has only one parent in level $l$.

Note that if all the refinement scores take their maximum values, then each node will have only one parent in the previous level, and the graph visualized will be a tree.
However, the parent-child relationships ($l' - l = 1$) do not have to be consistent with the ancestor-descendant ($l' - l > 1$) relationships for all of the refinement scores to be 1.

\subsection*{Minimizing $r(v)$}

Suppose we want to minimize $r(v)$ for a node $v$ in level $l$.
We can write the inner sum in the definition of $r(v)$ as
\begin{align*}
  \frac{1}{\sum_{w \in V_{l'}} w(v,w)} &\sum_{v'_{l'} \in V_{l'}} \frac{w(v, v'_{l'})^2}{\sum_{u \in V_l} w(u, v'_{l'})} \\
  &= \frac{1}{\sum_{w \in V_{l'}} w(v,w)} \sum_{v'_{l'} \in V_{l'}} \frac{w(v, v'_{l'})}{1 + \sum_{u \in V_l \setminus \{v\}} w(u, v'_{l'}) / w(v, v'_{l'})}
\end{align*}
Supposing that the weights $w(v, w)$, $w \in V_{l'}$ are fixed, the quantity above goes to zero when for each $v'_{l'}$ s.t. $w(v, v'_{l'})> 0$,  $w(u, v'_{l'}) \to \infty$ for some $u \in V_l \setminus \{v\}$. The refinement score $r(v)$ is an average over these values, and so the refinement score will also go to zero.
Therefore, for $r(v)$ to be small, all the descendants of $v$ need to primarily descend from some other node in the same level as $v$.

\subsection*{Refinement scores when all the weights are equal}

One of the ``edge'' cases we are particularly interested in is one in which all the weights are equal.
This is our intuition about what will happen if the clusters in the different levels don't correspond to each other at all.

If all the edges are equal, no matter what $l$ is, we will have $\win(v, v'_{l'}) = 1$, and so the expression for the refinement score simplifies to
\begin{align*}
  r(v) &= \frac{|V_l|}{L-l} \sum_{l' = l + 1}^L \sum_{v'_{l'} \in V_{l'}} \wout(v, v'_{l'}) \win(v, v'_{l'})\\
  &= \frac{|V_l|}{|V_l|(L-l)} \sum_{l' = l + 1}^L \sum_{v'_{l'} \in V_{l'}} \wout(v, v'_{l'}) = 1
\end{align*}

Overall, these results show us how the refinement scores work, and give us some insight into how the weights that we don't visualize enter into the refinement score calculations.
For example, if we had an alignment graph for which all the weights between subsequent levels were equal, we could still have a node with a relatively high refinement score if the weights that we didn't see satisfied the criteria for maximizing the refinement score (each node in a later level has only one ancestor in the level of the node we are interested in).
On the other hand, the alignment graph could look like a tree when we just look at the weights between subsequent levels, but if the weights in the levels we don't see are either all equal or such that the node we are interested in doesn't have descendants in the later levels, its refinement score could be very small.

\section{Comparing diagnostics}

The diagnostics measure different properties of an alignment. Both low coherence / high refinement and low refinement / high coherence combinations are possible, although in the examples below the diagnostics tend to track each other. We would expect the refinement score to be high but the coherence score to be low in the case that the alignment plot has a branching structure. If we do the calculations for an alignment (or piece of an alignment) as shown in Supplementary Figure \ref{fig:coherence_refinement_toy}, we can see that the refinement score for the highlighted node will be 1 (the largest possible value for that node), but the coherence score will be $\frac{1}{2}(p + p^2)$. Plugging in similar numbers, where each of the children has only a as an ancestor in that level, with more branching downstream of the highlighted node shows that as the branching below that node increases, the coherence score for a decreases but the refinement score stays at 1.

On the other hand, the refinement score can be small for a topic with high coherence if that topic doesn't have many descendants. As we see in the example in Supplementary Figure \ref{fig:coherence_refinement_toy}, the highlighted node has a coherence score of 1 because it is connected with weight 1 to the only other node in its same path. On the other hand, it has a low refinement score of $\frac{3  \delta}{1 + \delta}$ because all of the nodes in the subsequent level are much more closely aligned to the other competing topics.

Overall, the coherence score describes how ``good'' or ``trustworthy'' a topic is; topics with high coherence scores appear consistently across levels. This is true even if the refinement score is low — in that case, the refinement score is likely to be low simply because the topic is present at low frequency. On the other hand, the combination of high refinement and low coherence score suggests that the topic is a mixture of several high-coherence topics. These topics can still be useful to the analyst, as they simply represent a coarser-grained summary of the data.

\begin{figure}
    \centering
    \includegraphics[height=.25\textwidth]{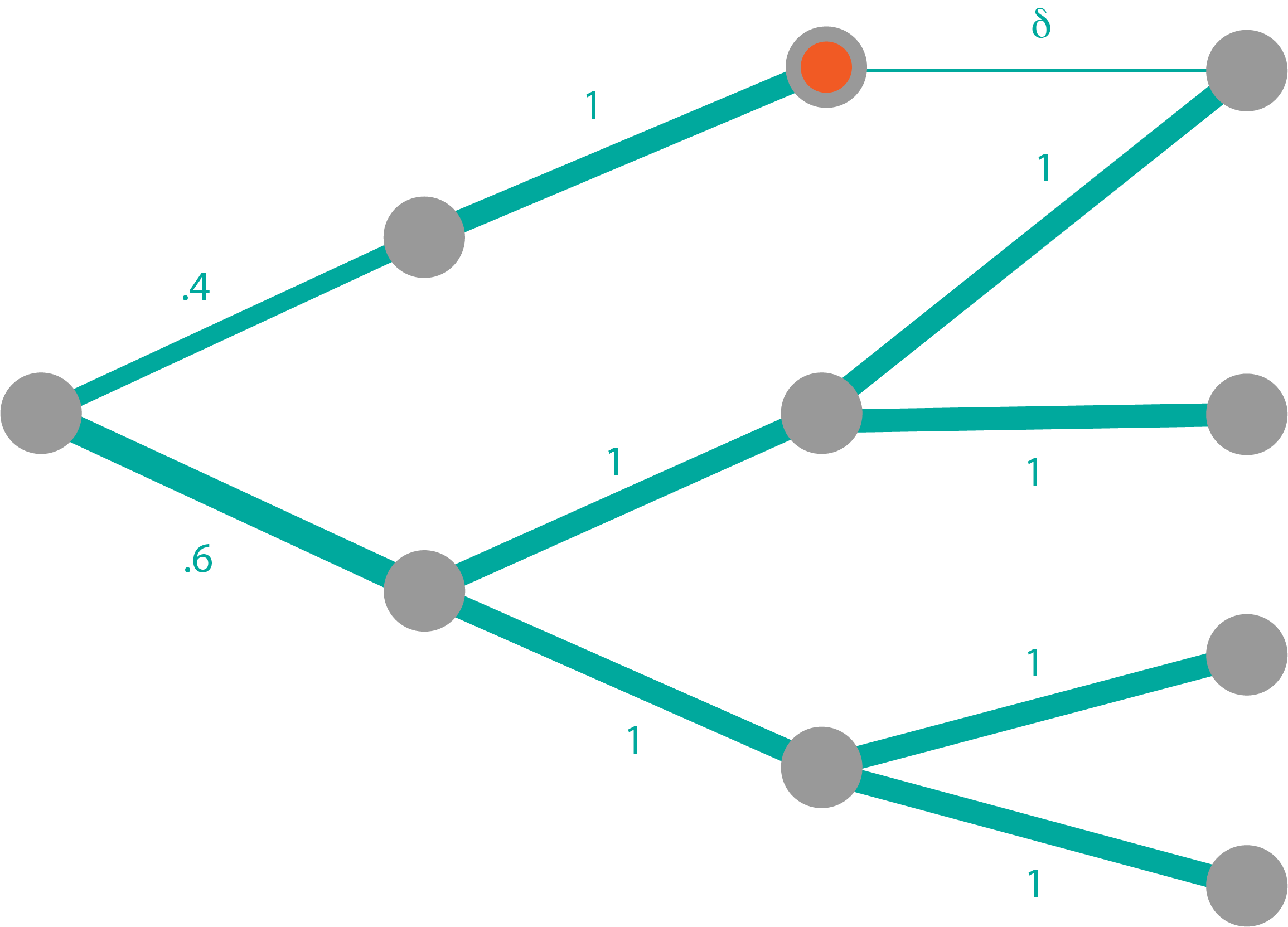}
    \hspace{.1\textwidth}
    \includegraphics[height=.25\textwidth]{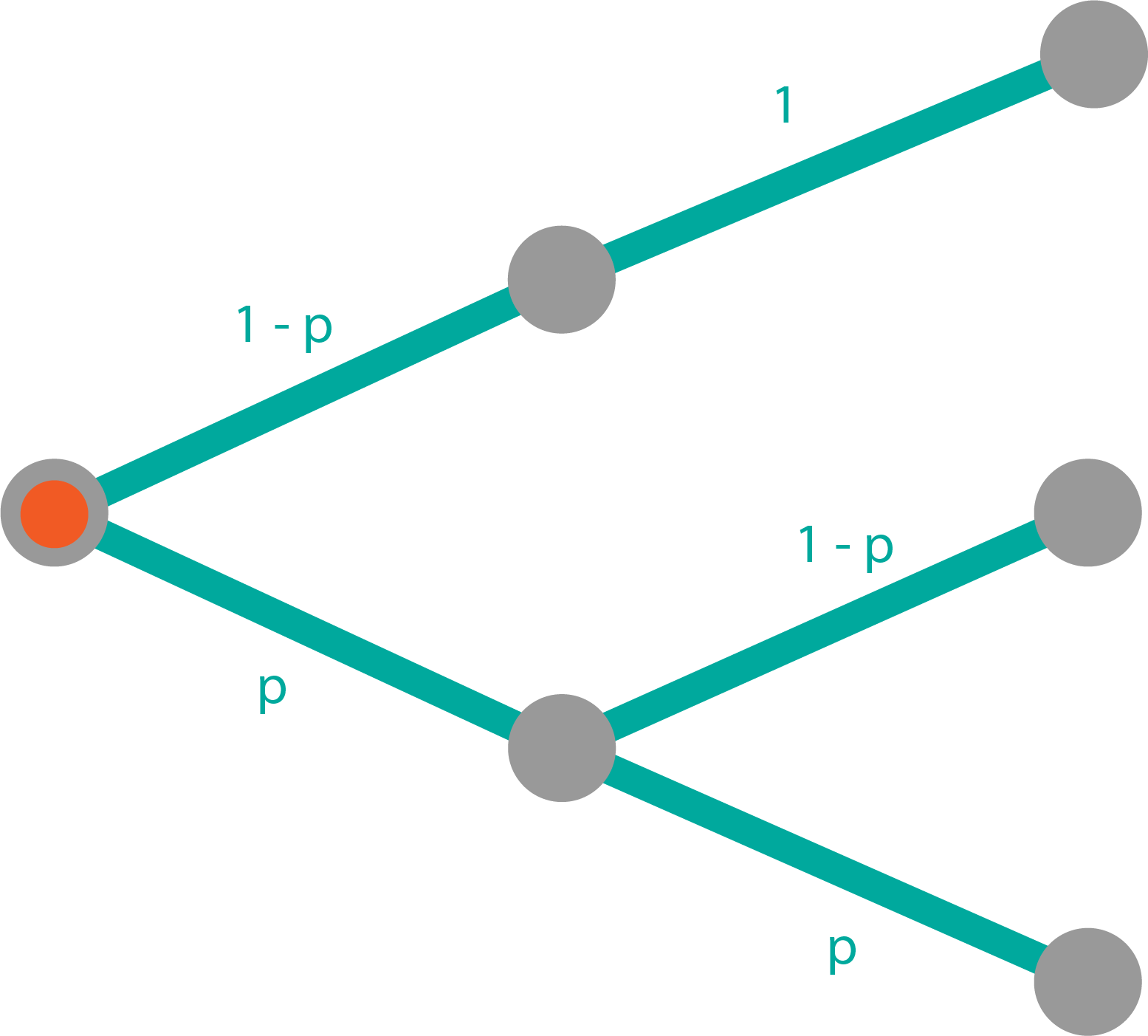}
    \caption{Examples of situations where coherence and refinement scores are not aligned. In the first example, the coherence score for the highlighted node is 1 but the refinement score is $3 \delta / (1 + \delta)$. In the second example, the refinement score for the highlighted node is 1 but the coherence score is $\frac{1}{2} (p + p^2)$.}
    \label{fig:coherence_refinement_toy}
\end{figure}

\newpage

\section{Comparing alignments of LDA-generated \textit{vs} null model datasets}

\begin{figure}[H]
    \centering
    \includegraphics[width=\textwidth]{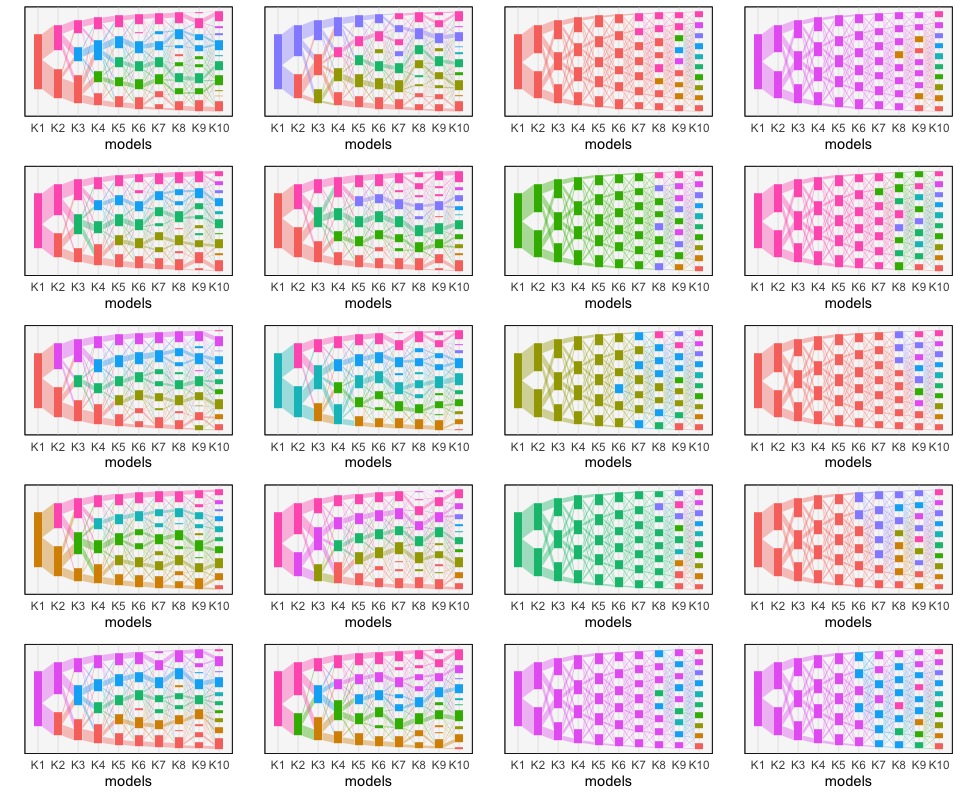}
    \caption{Product alignment colored by path for simulated data. Left two columns correspond to data coming from the true LDA model with $K = 5$, and right two columns correspond to data coming from a null model.}
    \label{fig:product_paths}
\end{figure}
\begin{figure}[H]
    \centering
       \includegraphics[width=\textwidth]{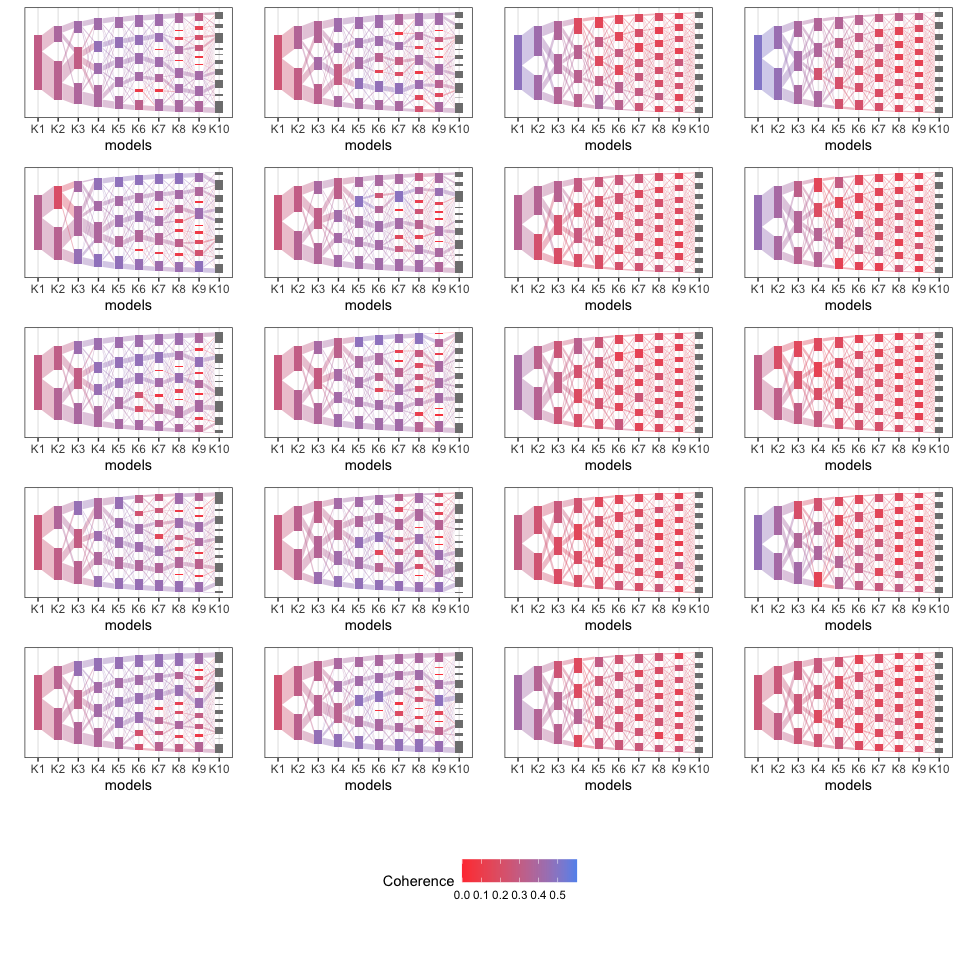}
    \caption{Product alignment colored by coherence score for simulated data. Left two columns correspond to data coming from the true LDA model with $K = 5$, and right two columns correspond to data coming from a null model.}
    \label{fig:product_coherence}
\end{figure}
\begin{figure}[H]
    \centering
     \includegraphics[width=\textwidth]{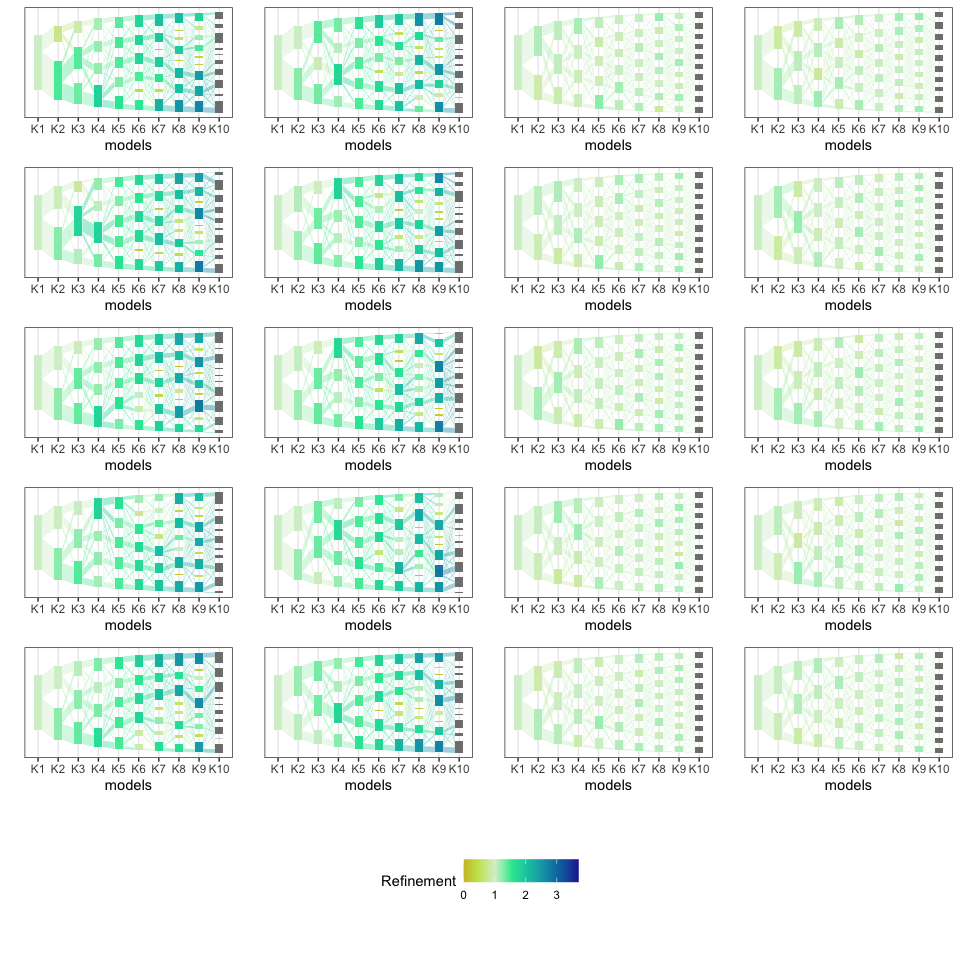}
        \caption{Product alignment colored by refinement score for simulated data. Left two columns correspond to data coming from the true LDA model with $K = 5$, and right two columns correspond to data coming from a null model.}
    \label{fig:product_refinement}
\end{figure}

\begin{figure}[H]
    \centering
    \includegraphics[width=\textwidth]{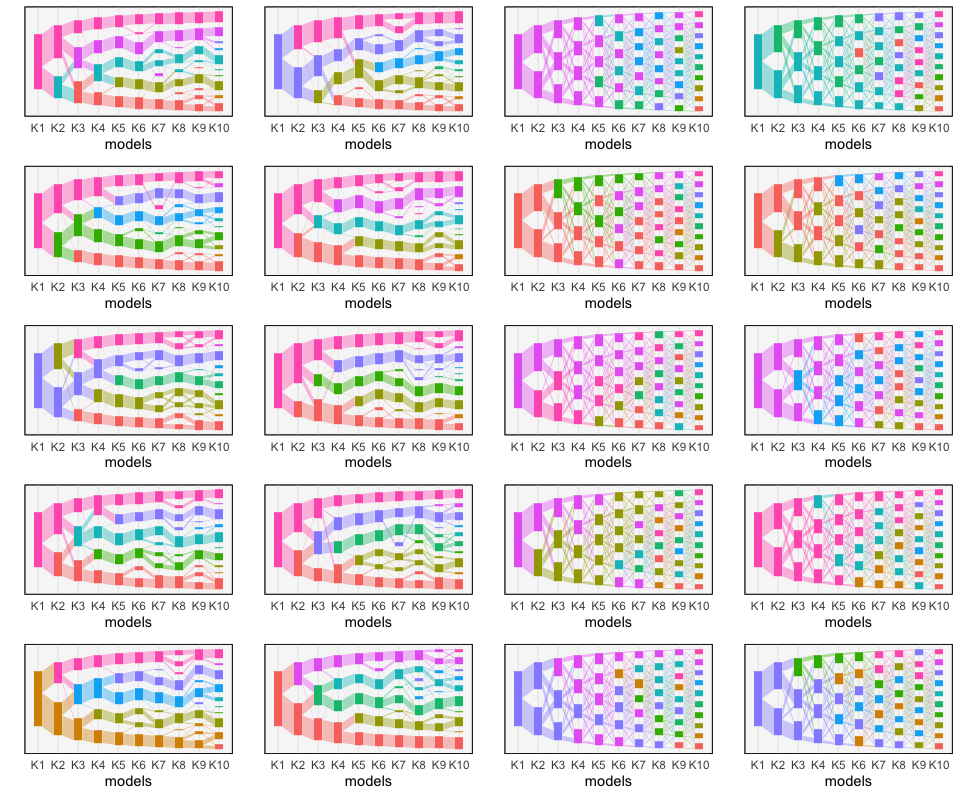}
    \caption{Transport alignment colored by path for simulated data. Left two columns correspond to data coming from the true LDA model with $K = 5$, and right two columns correspond to data coming from a null model.}
    \label{fig:transport_paths}
\end{figure}

\begin{figure}[H]
    \centering
       \includegraphics[width=\textwidth]{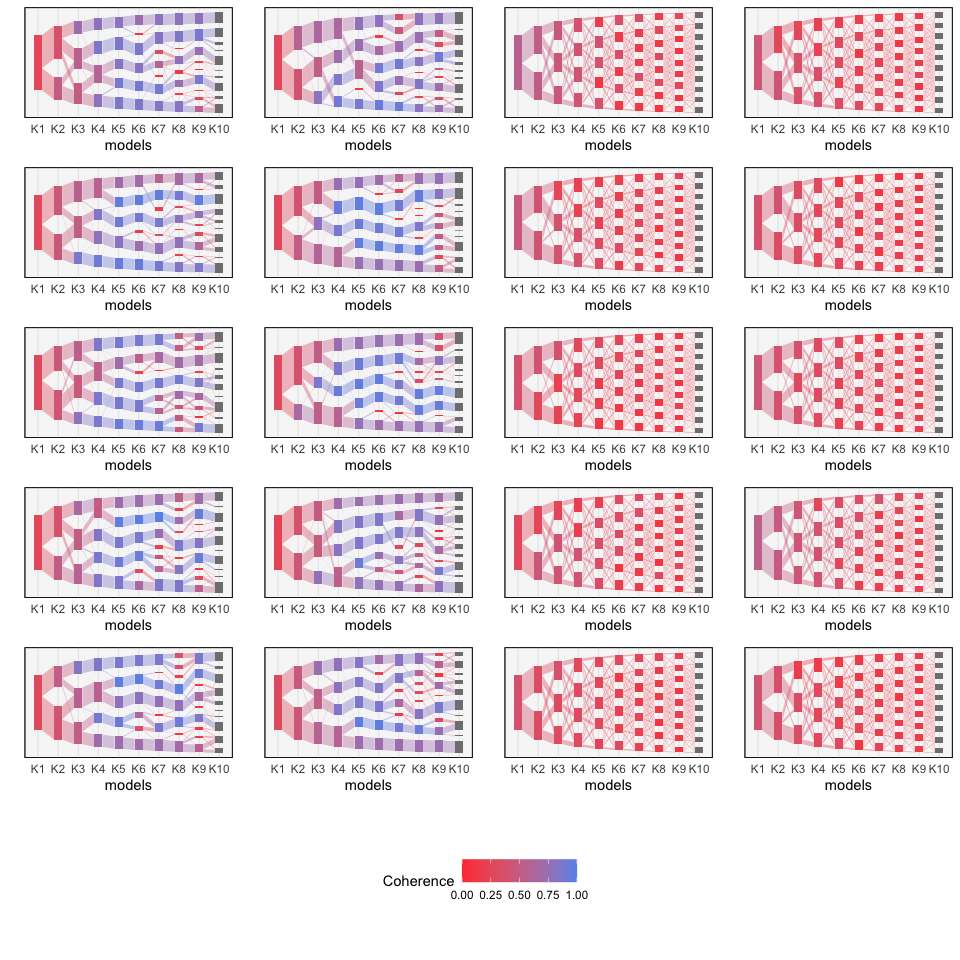}
    \caption{Transport alignment colored by coherence score for simulated data. Left two columns correspond to data coming from the true LDA model with $K = 5$, and right two columns correspond to data coming from a null model.}
    \label{fig:transport_coherence}
\end{figure}
\begin{figure}[H]
    \centering
     \includegraphics[width=\textwidth]{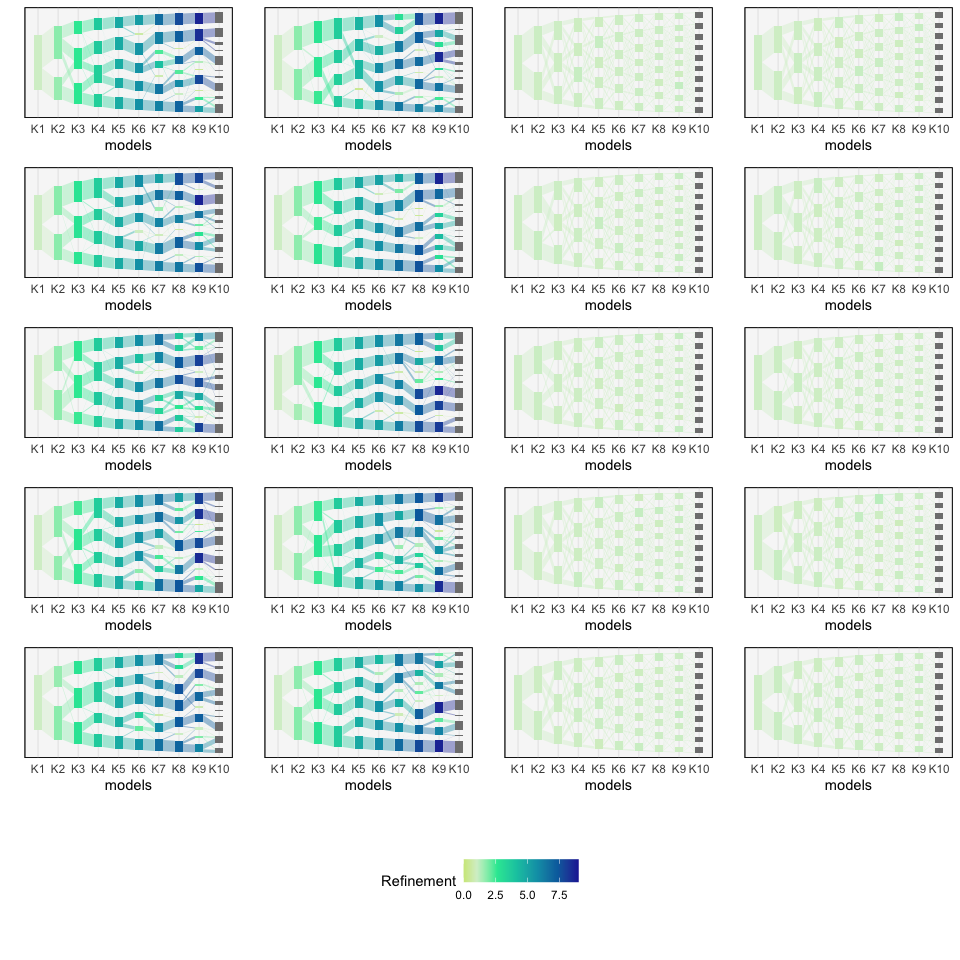}
        \caption{Transport alignment colored by refinement score for simulated data. Left two columns correspond to data coming from the true LDA model with $K = 5$, and right two columns correspond to data coming from a null model.}
    \label{fig:transport_refinement}
\end{figure}

\newpage

\section{Example of alignments with increasing level of background noise}

\begin{figure}[H]
    \centering
    \includegraphics[width=0.48\textwidth]{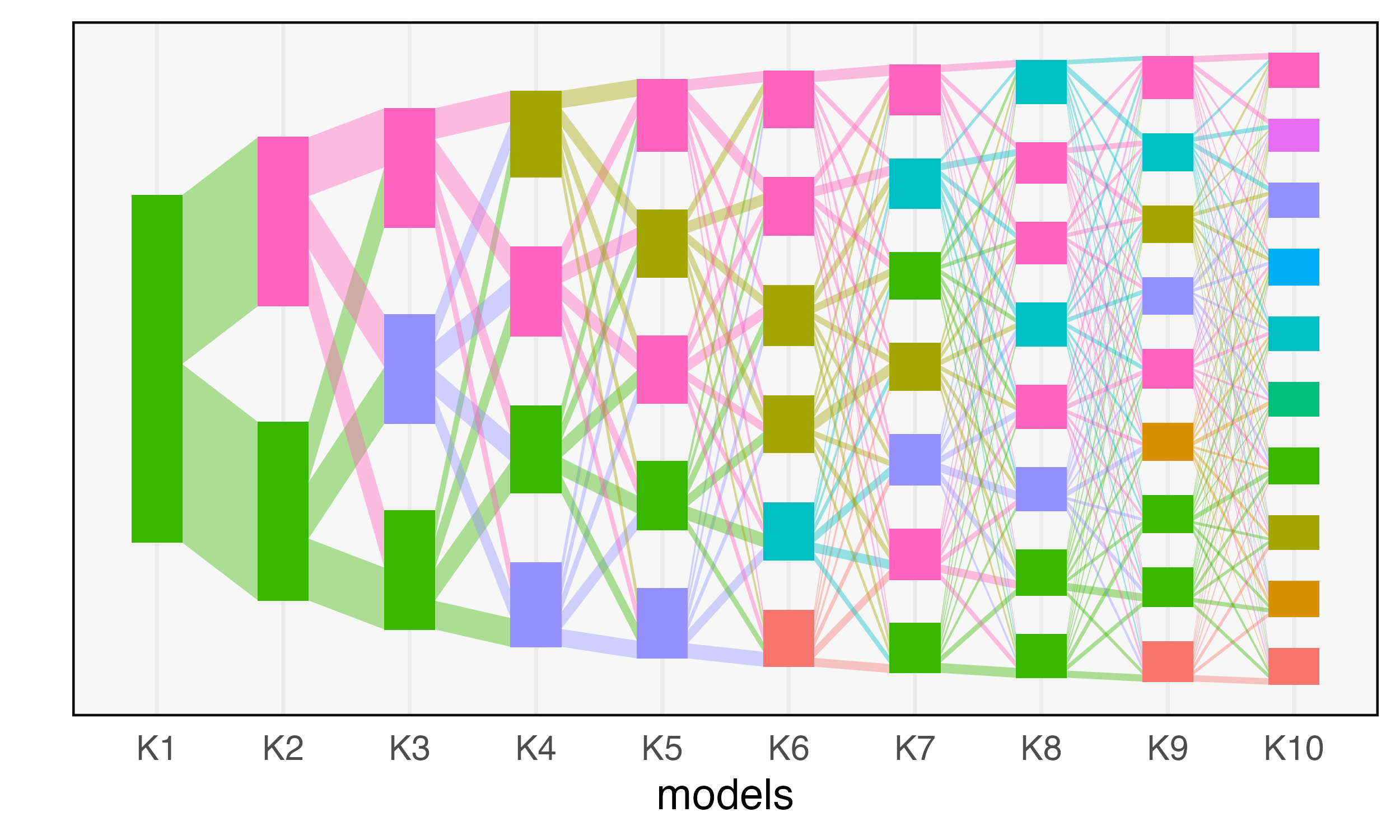}
    \includegraphics[width=0.48\textwidth]{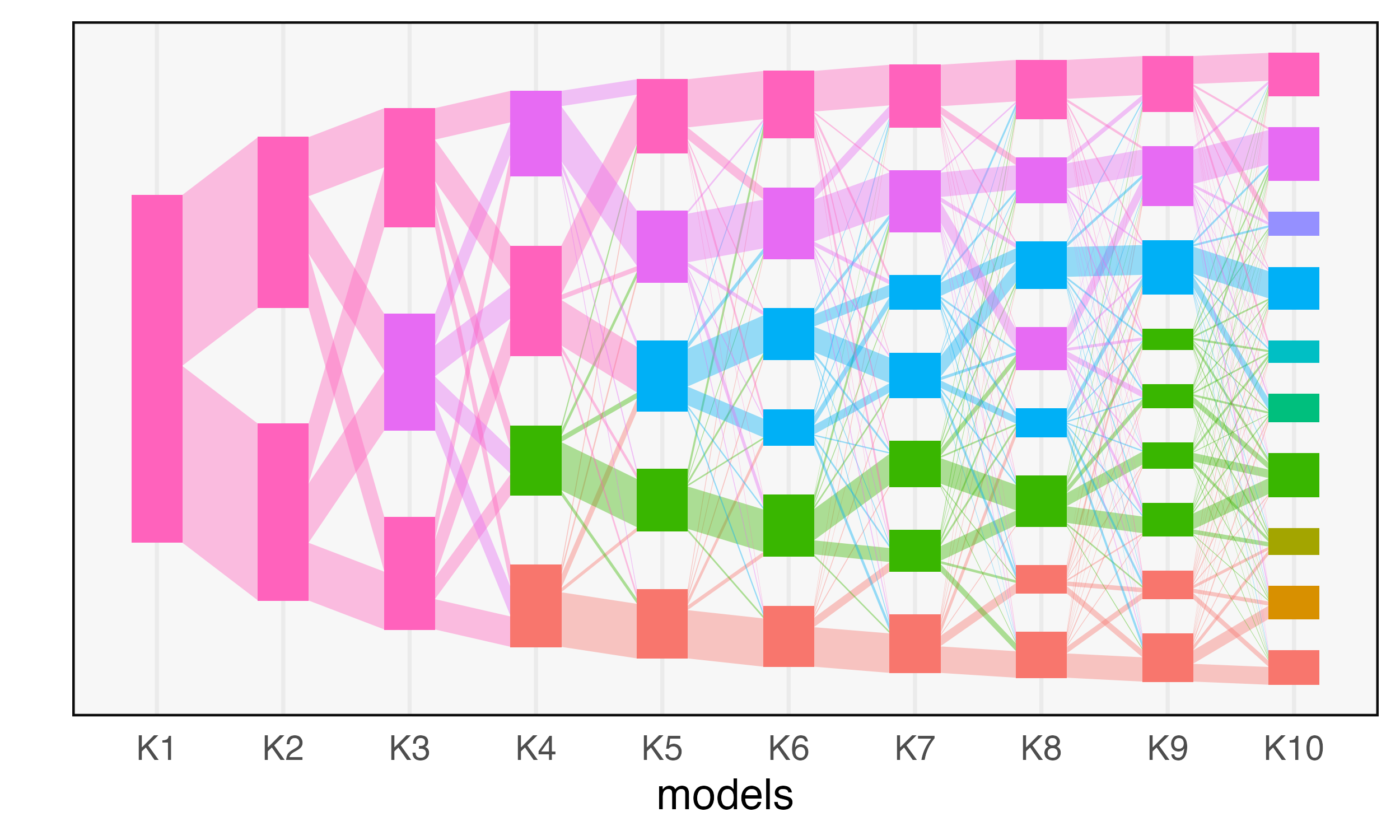}
    \includegraphics[width=0.48\textwidth]{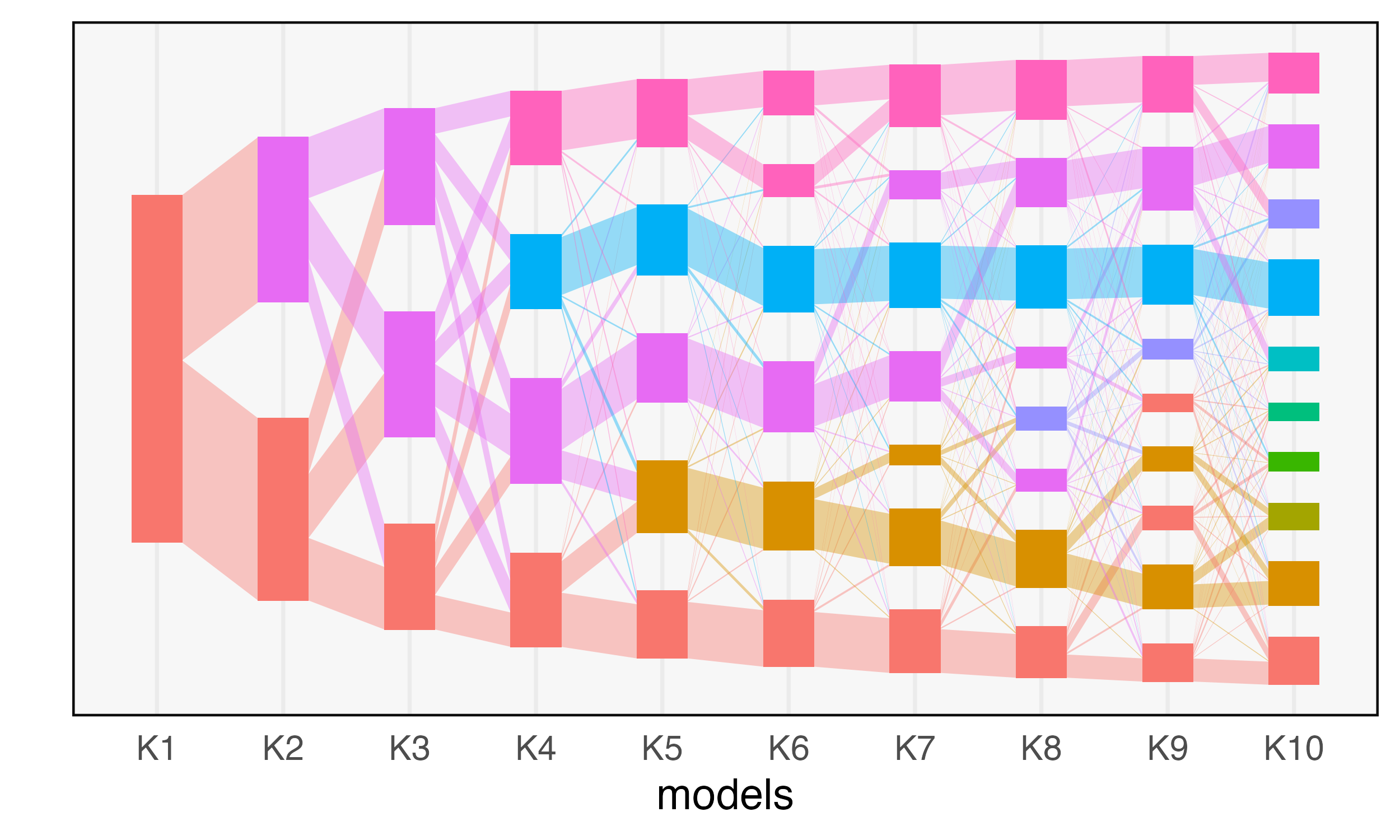}
    \includegraphics[width=0.48\textwidth]{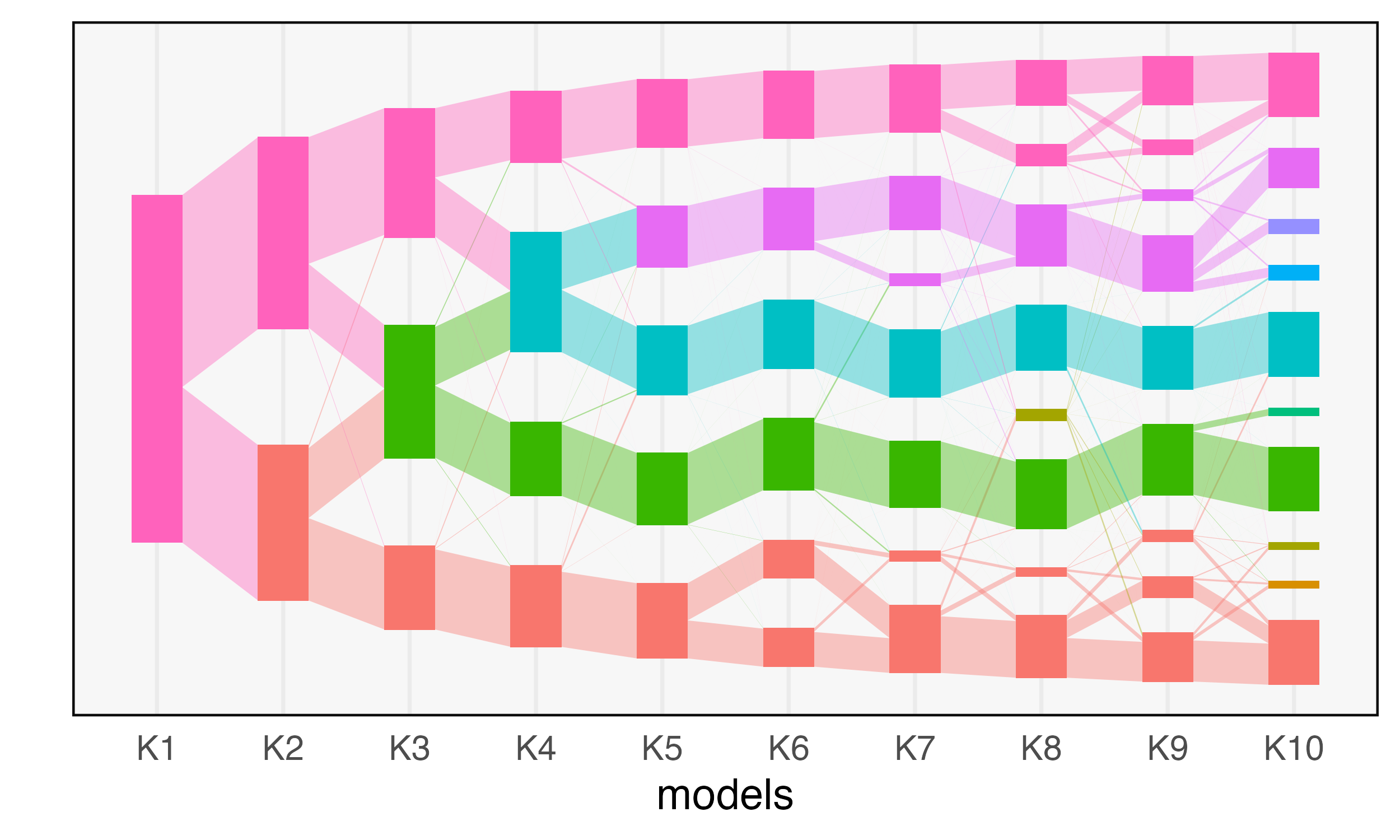}
    \caption{Flows for LDA with background variation at levels $\alpha \in \{0, 0.4, 0.6, 1\}$. A more definitive topic structure emerges for larger $\alpha$, with less exchange between neighboring branches.}
    \label{fig:lda_flow_gradients}
\end{figure}

\newpage

\section{Convergence of diagnostics as $N$ increases}

\begin{figure}[H]
    \centering
     \includegraphics[width=\textwidth]{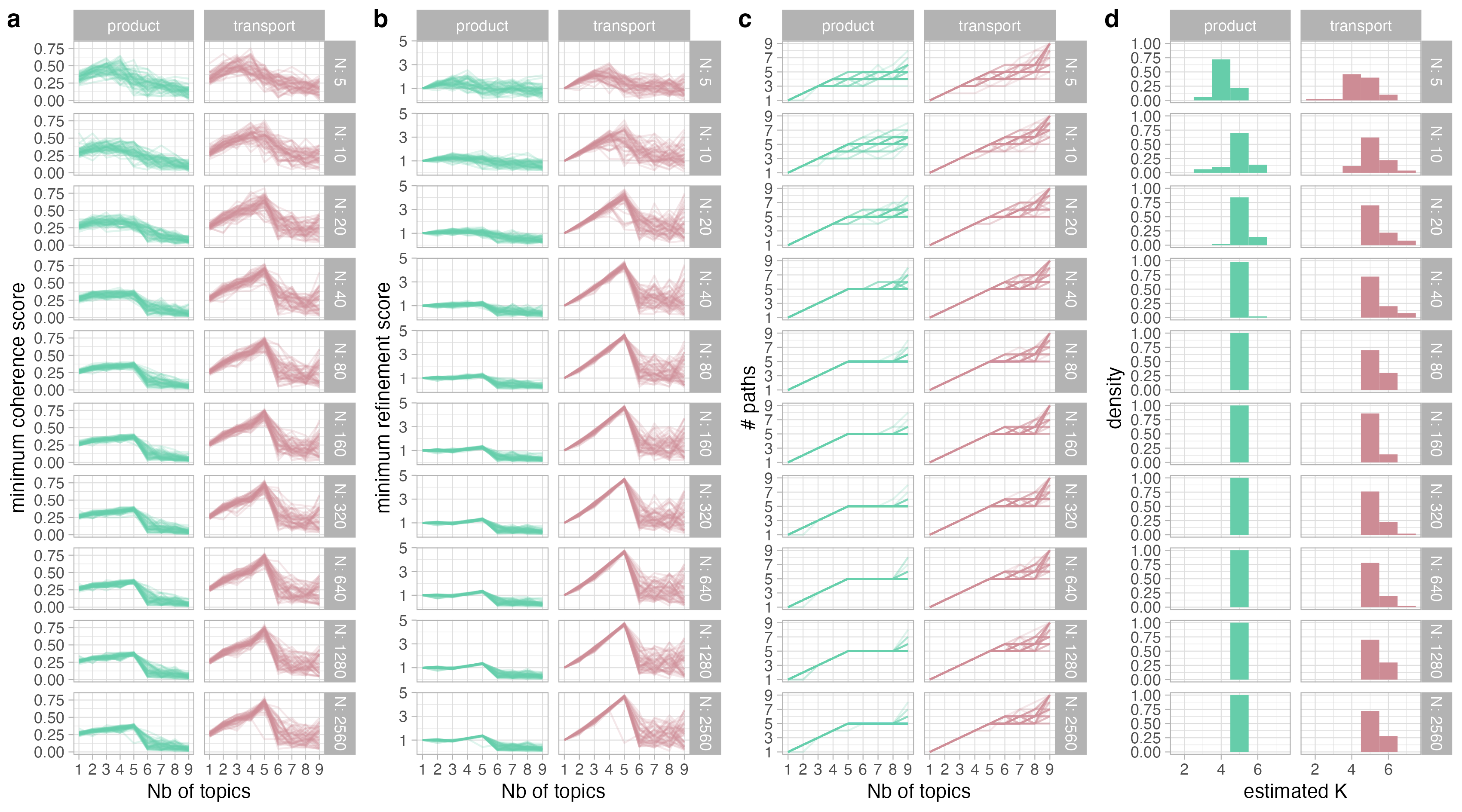}
        \caption{Summary scores as the number of samples ($N$) in simulated datasets increases (vertical panels). For each $N$, 50 datasets were generated and topic alignment was performed on each dataset. Each line represents the score summary for one dataset. Panel (a) and (b) show the minimum of these scores for each simulated dataset and model. We chose to show the minimum of the scores because we observed in simulations that "spurious" topics introduced at higher resolution were characterized by low coherence and/or refinement scores. Consequently, the minimum of the scores allows to identify drop-offs in the lower envelope for the scores. Panel (c) shows the number of path identified at each resolution. Panel (d) shows the distribution of the number of path at which a plateau is identified in panel (c).}
    \label{fig:asymptotic_behavior}
\end{figure}

\newpage

\section{Strain switching across $S$}

In this appendix, we extend the discussion of strain switching. We provide details of the perturbation mechanism (Algorithm \ref{alg:perturbations}) and investigate the sensitivity of topic alignment across a wider range of $S$.
 We simulate strain-switching data for $S \in \{10, 30, \dots, 230\}$. For each choice of $S$, we generate 50 datasets and align topic models across a range $K = 2, \dots, 10$ of topics.

\begin{figure}[H]
    \centering
    \includegraphics[width=0.9\textwidth]{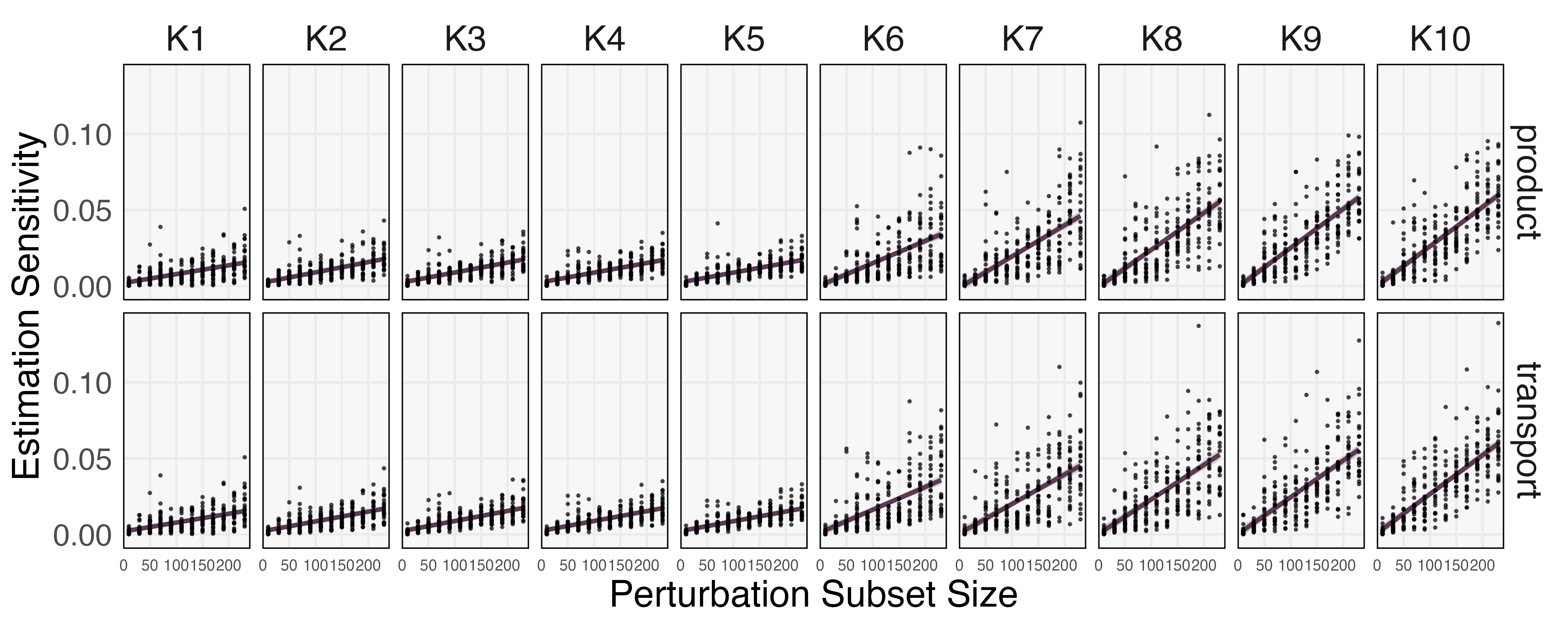}
    \caption{The ability of models to detect strain switching as a function of $K$ and $S$. Model resolution increases across panels moving from left to right. Within each panel, the size of the number of swapped species $S$ is plotted against the estimation sensitivities defined in the main text. The larger the subset $S$, the higher the sensitivity. Further, high-resolution models can more easily distinguish perturbed topics, as indicated by the steeper slopes for panels on the right.}
    \label{fig:equivalence_similarities}
\end{figure}

To gauge sensitivity to perturbed topics, we measure cosine similarities 
across simulation replicates. If strain switching cannot be detected, then we expect $\xi_{1k}^{m} \approx \xi_{2k}^{m}$ and $\xi_{3k}^{m} \approx \xi_{4k}^{m}$ for all $k, m$ — the estimated topics will lack specificity for any member of the equivalent pairs. Figure \ref{fig:equivalence_similarities} shows the estimation specificity, $\frac{1}{K}\sum_{k = 1}^{m}\absarg{\xi_{1k}^m - \xi_{2k}^m} + \absarg{\xi_{3k}^m - \xi_{4k}^m}$ for each of the 50 replicates for each $S$. This statistic quantifies the difference between rows 1-2 and 3-4 visible in the heatmap of topic similarities, but across all simulation replicates.

As expected, larger perturbations are more easily detected. For models with $K \leq 5$, there is a small increase in estimation specificity as $S$ increases; strain switching might have a small effect on the dominant signatures in the data. For $K > 5$, the specificity as a function of $S$ steepens -- more highly resolved topics can more easily distinguish between perturbed topics.

\begin{algorithm}
 \KwData{Topics $\beta_{k}$, subset size $S$ to perturb, $\tilde{K}$ of the topics to perturb, number of perturbations $R$.}
 \For{$k \leq \tilde{K}$}{
 Sample $S$ coordinates to perturb, and define a mapping $\pi$ such that $\pi\left(s\right)$ provides the $s^{th}$ perturbed index. \\
 \For{$r \leq R$}{
 For the subset $S$, draw $\nu_{k}^{r} \sim \Dir\left(\lambda_{S} 1_{\absarg{S}}\right)$\\
 Renormalize $\nu_{k}^{r} := \frac{\|\beta_{k}\left[S\right]\|_{1}}{\|\nu_{r}\|_{1}}\nu_{r}$ \\
 Perturb $\beta_{k}$ at coordinates specified by $S$,
 \begin{align*}
\tilde{\beta}_{kd}^{r} := \begin{cases} \beta_{kd} & \text {if } d \notin S \\
\nu_{k\pi\left(s\right)}^{r} & \text {otherwise.}
\end{cases}
\end{align*}
 }
 }
 \caption{Strategy for generating perturbed topics.}
 \label{alg:perturbations}
\end{algorithm}

\newpage

\section{Perplexity comparisons} 
\label{sec:perplexity}

Perplexity is defined as 
\begin{align}
\label{eq:perplexity}
\text{perplexity}\left(x_{1}^{\ast}, \dots, x_{n}^{\ast}\right) = \exp{-\frac{\sum_{i = 1}^{n} \log p\left(x^{\ast}_{i}\right)}{\sum_{i = 1}^{n} N_i}},
\end{align}
where $N_i$ is the total count of document $i$. 
Hence, test documents $x_{i}^{\ast}$ with low likelihood-per-read 
under the fitted model $p$ have high perplexity.

\begin{figure}[H]
    \centering
    \includegraphics[width=0.8\textwidth]{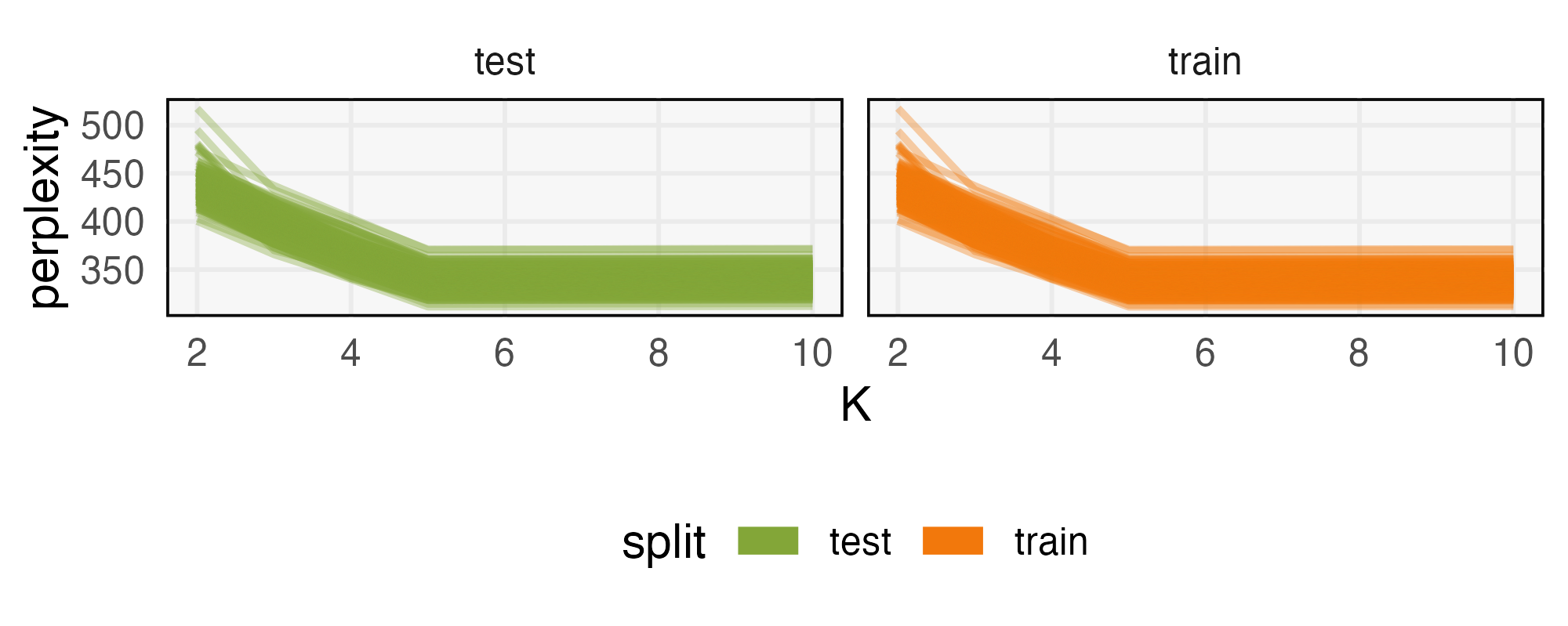}
    \caption{Perplexity for train and test samples for data generated by a true LDA model with $K = 5$ topics. The ``elbow'' in train and test perplexity can be used to detect the true value of $K$.}
    \label{fig:lda-perplexity}
\end{figure}

\begin{figure}[H]
    \centering
    \includegraphics[width=0.9\textwidth]{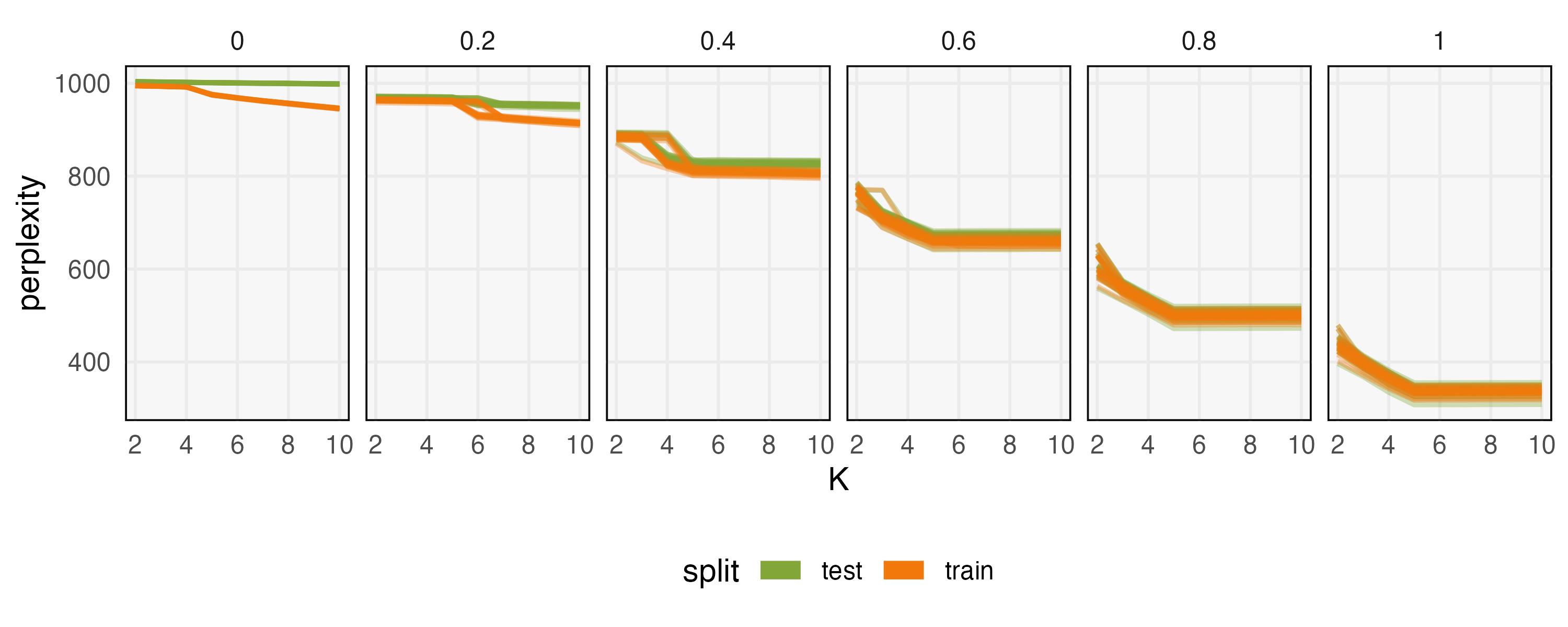}
    \caption{Perplexity for train and test samples from data generated by an LDA model with varying levels of background noise. Panel titles match the $\alpha$ from the corresponding simulation in the main text. For smaller $\alpha$, the ``elbow'' in perplexity sometimes appears at incorrect values of $K$ (e.g., 6 - 7 for $\alpha = 0.2$ and 4 - 5 for $\alpha = 0.4$). The specific locations of these drop-offs is dependent on the $\lambda_{\nu}$ hyperparameter generating this background noise.}
    \label{fig:gradient-perplexity}
\end{figure}

\begin{figure}
    \centering
    \includegraphics[width=0.4\textwidth]{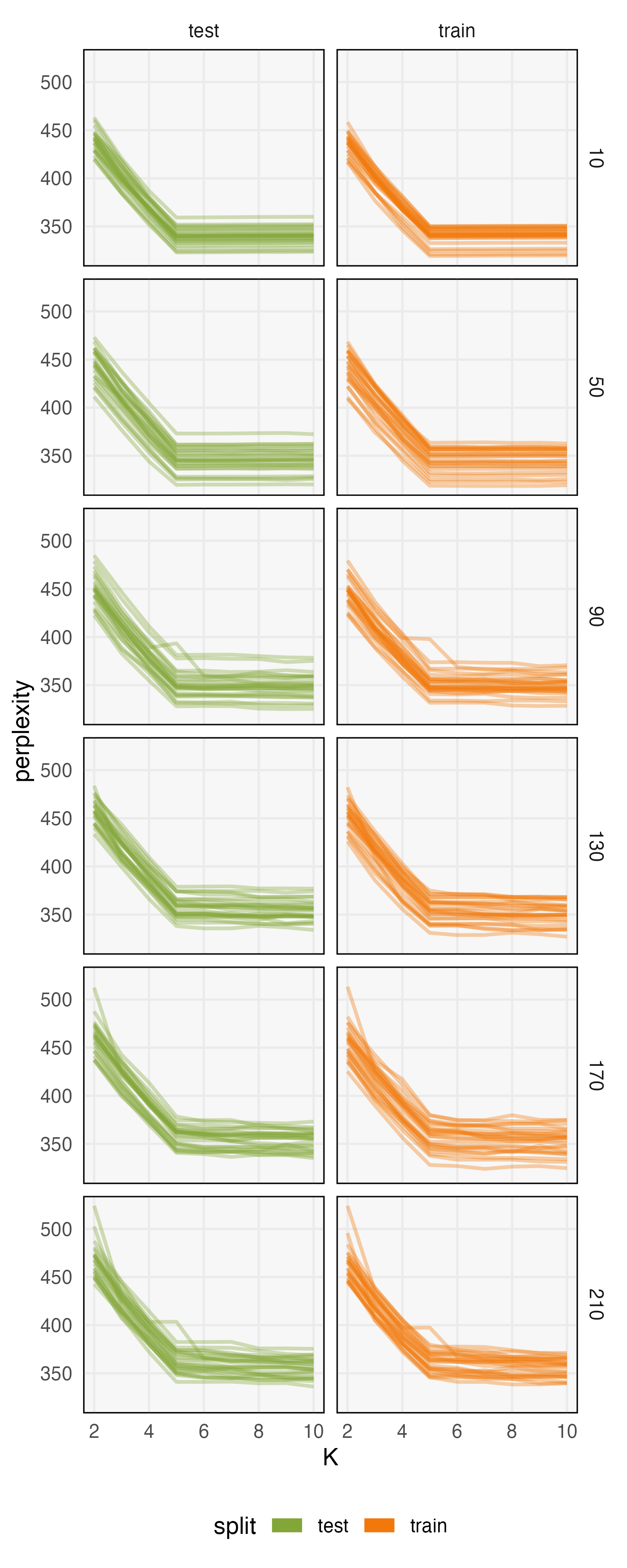}
    \caption{In and out-of-sample perplexity for data generated according to the strain switching setup. Rows provide different values of $S$, the number of switched strains. Perplexity at small $K$ is slightly larger when $S$ is large. Further, for large $S$, perplexity continues to decrease slightly even beyond the ``elbow'' at $K = 5$. However, no structure at $K = 7$ suggests that two of the topics may exhibit switching behavior.}
    \label{fig:switching-perplexity}
\end{figure}

\newpage

\section{Hierarchical LDA comparison}

\begin{figure}[H]
    \centering
     \includegraphics[width=\textwidth]{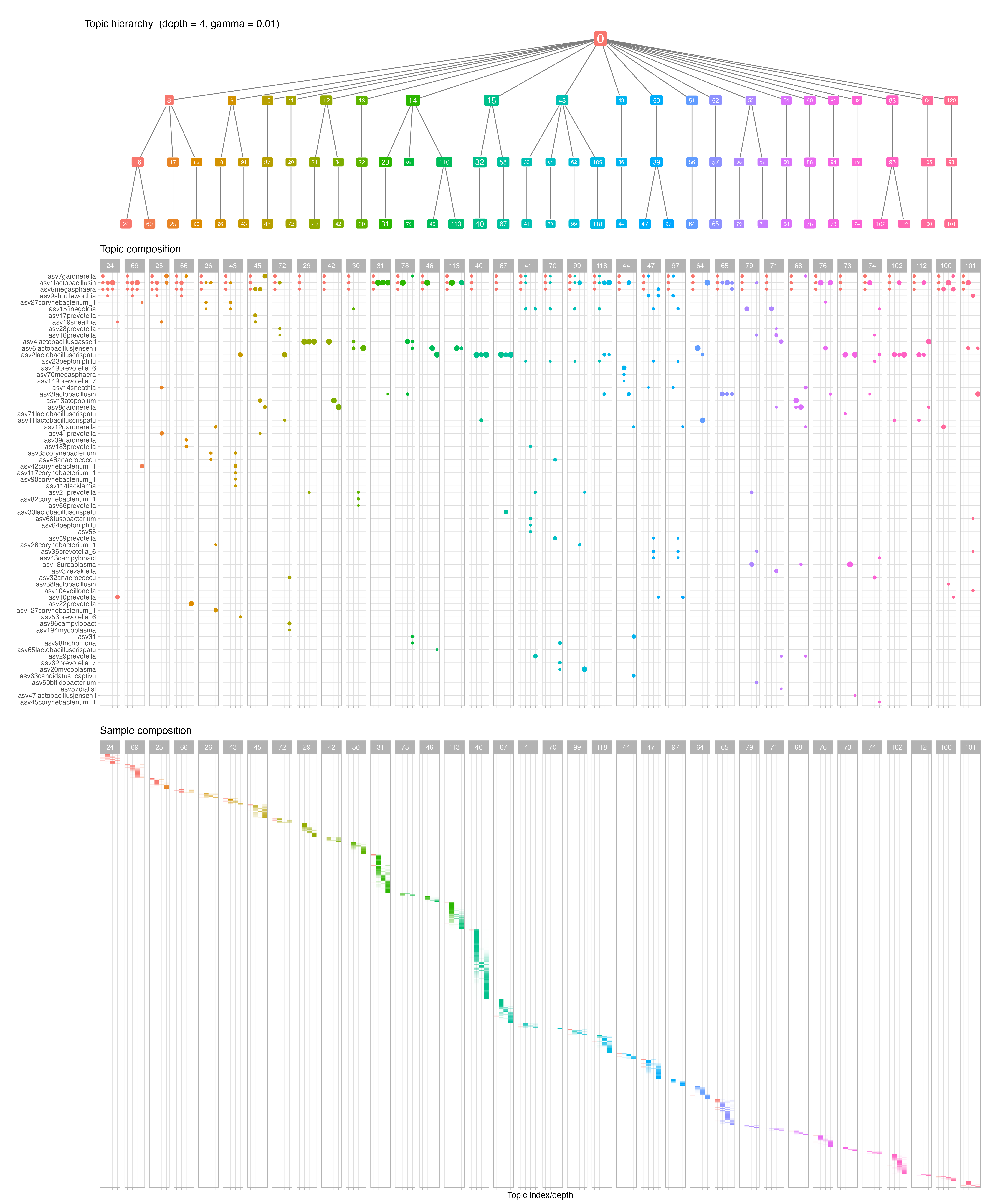}
     \caption{Hierarchical LDA (hLDA) model of the vaginal microbiome data.  }
     \vspace{-2em}
\flushright \emph{(continued)\\[.25em]}
\hrule
\end{figure}

\begin{figure}[t!]
  \ContinuedFloat
  \caption{ (Caption continued.)
  hLDA was fit to a subset of the vaginal microbiome data with a depth of 4 and a concentration parameter (gamma) of 0.1. (Top) Hierarchical structure of hLDA topics.  
    (Middle) Topic composition for each vertical path of the hierarchical structure. There is a vertical path for each leave of the hierarchical tree and path are labeled and colored according to the leave topic number. Each path is shown on a vertical panels. Topics from each vertical path are on the x-axis, ordered by depth, starting with the root topic on the left (This implies that the topic composition of the root topic is repeated for each vertical path). Features ("words") are shown vertically. The dot size is proportional to the proportion of a feature in a topic. The dot color is set to match the color of the topics on the top panel. (Bottom) Sample composition in terms of topic proportion. Horizontal panel and x-axis are the same as in the middle panel. Each horizontal line represents a sample. Given that samples can only be composed of topics on a given vertical path, samples have been assigned to their path and ordered by path. Colors match the topics color on the top panel. Transparency is inversely proportional to topic proportion in each sample.}
  \hrule
\end{figure}

\end{document}